\documentclass[prl,twocolumn,floatfix,superscriptaddress,longbibliography]{revtex4-1}
\usepackage{amssymb,amsmath,amstext}                
\usepackage{graphicx}                                               
\usepackage{epstopdf}                                               
\usepackage{color}       
\usepackage{bm}
\usepackage{appendix}                                              
\usepackage[utf8]{inputenc}
\usepackage{bbold}
\usepackage{bbm}
\usepackage{latexsym}
\usepackage{dsfont}
\usepackage[colorlinks=true,citecolor=blue,linkcolor=magenta]{hyperref}
\usepackage{newfloat,algcompatible}

\DeclareFloatingEnvironment[
    fileext=loa,
    listname=List of Algorithms,
    name=Pseudocode,
    placement=tbhp,
]{algorithm}

\global\long\def\avg#1{\langle#1\rangle}

\global\long\def\ket#1{|#1\rangle}

\global\long\def\tr{\mathrm{tr}}
\global\long\def\dg{\dagger}
\global\long\def\im{\imath}

\global\long\def\l{\mathcal{L}}
\global\long\def\r{\mathcal{R}}
\global\long\def\s{\mathcal{S}}
\global\long\def\i{\mathcal{I}}
\global\long\def\u{\mathcal{U}}

\global\long\def\c{\mathcal{C}}

\begin{document}
\title{Breaking the entanglement barrier: Tensor network simulation of quantum transport}

\author{Marek M. Rams}
\email{marek.rams@uj.edu.pl}
\affiliation{Jagiellonian University, Marian Smoluchowski Institute of Physics, \L{}ojasiewicza 11, 30-348 Krak\'ow, Poland}

\author{Michael Zwolak}
\email{mpz@nist.gov}
\affiliation{Biophysics Group, Microsystems and Nanotechnology Division, Physical
Measurement Laboratory, National Institute of Standards and Technology,
Gaithersburg, MD 20899, USA}

\begin{abstract}
    The recognition that large classes of quantum many-body systems have limited entanglement in the ground and low-lying excited states led to dramatic advances in their numerical simulation via so-called tensor networks. However, global dynamics elevates many particles into excited states, and can lead to macroscopic entanglement and the failure of tensor networks. Here, we show that for quantum transport -- one of the most important cases of this failure -- the fundamental issue is the canonical basis in which the scenario is cast: When particles flow through an interface, they scatter, generating a ``bit'' of entanglement between spatial regions with each event. The frequency basis naturally captures that -- in the long--time limit and in the absence of inelastic scattering -- particles tend to flow from a state with one frequency to a state of identical frequency. Recognizing this natural structure yields a striking -- potentially exponential in some cases -- increase in simulation efficiency, greatly extending the attainable spatial- and time-scales, and broadening the scope of tensor network simulation to hitherto inaccessible classes of non-equilibrium many-body problems.
\end{abstract}

\maketitle


Tensor networks enable the systematic search for ground states of certain many-body Hamiltonians, as well as numerical time evolution, provided that there is a limited amount of entanglement present~\cite{orus_tensor_2018,ran_review_2017,orus_practical_2014,eisert_entanglement_2013,schollwock_density-matrix_2011,verstraete_review_2008}. Quantum quenches -- when a parameter of the Hamiltonian is suddenly changed -- can, though, generate highly-entangled states, seen both experimentally~\cite{kaufman_quantum_2016} and theoretically~\cite{alba_entanglement_2017,liu_entanglement_2014,kim_ballistic_2013,schachenmayer_entanglement_2013,schuch_entropy_2008,calabrese_evolution_2005}. The large amount of entanglement creates a challenge for tensor network simulation and the efficient representation of the underlying quantum state~\cite{schuch_entropy_2008,eisert_entanglement_2013}. There are many approximate approaches under development to truncate further the description of the state and maintain control over the size of the tensor network \cite{Refael_2018, Altman2017, Luca_2018}, but these rely on additional assumptions, such as the thermalizing nature of the dynamics. We will here develop a controllable approach to break the entanglement barrier for an important class of problems in transport.

\begin{figure}[t!]
\begin{centering}
\includegraphics[width=1\columnwidth]{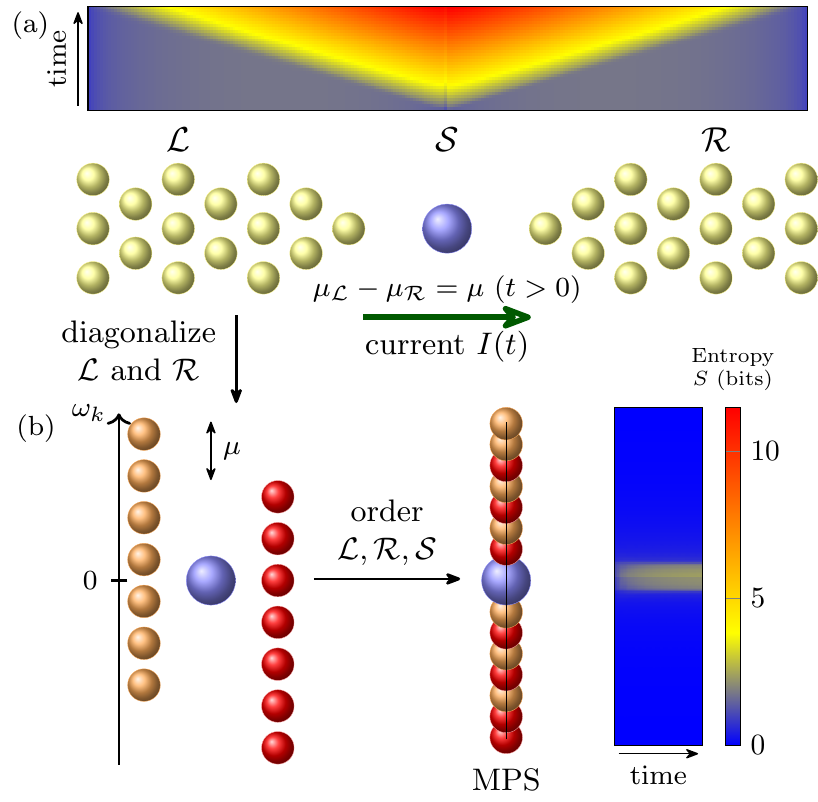}
\par\end{centering}
\caption{{\bf Entanglement, transport, and simulation.} {(a)} A bias or particle imbalance between the $\l$ and $\r$ reservoirs drives a current $I$ through the impurity system $\s$. This spatial structure, though, has an entanglement ``light cone'' (over top heat map), leading to macroscopic entanglement and simulation failure. {(b)} Separately diagonalizing the single-particle eigenstates in the $\l$ and $\r$ regions (of arbitrary spatial dimension) and combining them into a joint $\l\r$ environment circumvents this issue by naturally structuring the simulated system. The entanglement then becomes localized within the bias window and highly suppressed (right heat map). The heat map scale is for both entanglement plots and the simulation details are the same as in Fig.~\ref{fig:Inonint}.
\label{fig:Schematic}}
\end{figure}

Quantum transport through an impurity $\s$ is a paradigmatic example of a ``pathological'' quench.  A global bias $\mu$ drives particles through an interface where they scatter, see Fig.~\ref{fig:Schematic}(a). For particles around the Fermi level ($\omega = 0$), each scattering event gives rise to an entangled electron-hole pair~\cite{beenakker_electron-hole_2006,klich_quantum_2009}
\begin{equation}
\sqrt{T(0)} \ket{0_\l 1_\r}+\sqrt{1-T(0)} \ket{1_\l 0_\r} \label{eq:EntProd}
\end{equation}
across the left ($\l$) and right ($\r$) reservoir regions. The left component of the state represents a particle transmitted from $\l$ to $\r$ with transmission probability $T(0)$ and the right component the reflected particle (the phase is unimportant here). Given that the attempt frequency is $\mu/2\pi$~\cite{levitov_charge_1993}, the entanglement entropy $S$ increases as 
\begin{equation}
S \approx H[T(0)] \frac{\mu t}{2 \pi}, \label{eq:EntvsTime}
\end{equation}
where $H[T(0)]$ is the binary entropy of the transmission probability and $t$ the time~\cite{beenakker_electron-hole_2006,klich_quantum_2009}. 
This linear growth of spatial entanglement -- and its ``light cone'' spread~\cite{chien_landauer_2014} (see the heat map in Fig.~\ref{fig:Schematic}(a) -- is due to the linear increase in the number of entangled electron-hole pairs, as expressed by Eqs.~(\ref{eq:EntProd},\ref{eq:EntvsTime}).

\begin{figure}[t]
\begin{centering}
\includegraphics[width=1\columnwidth]{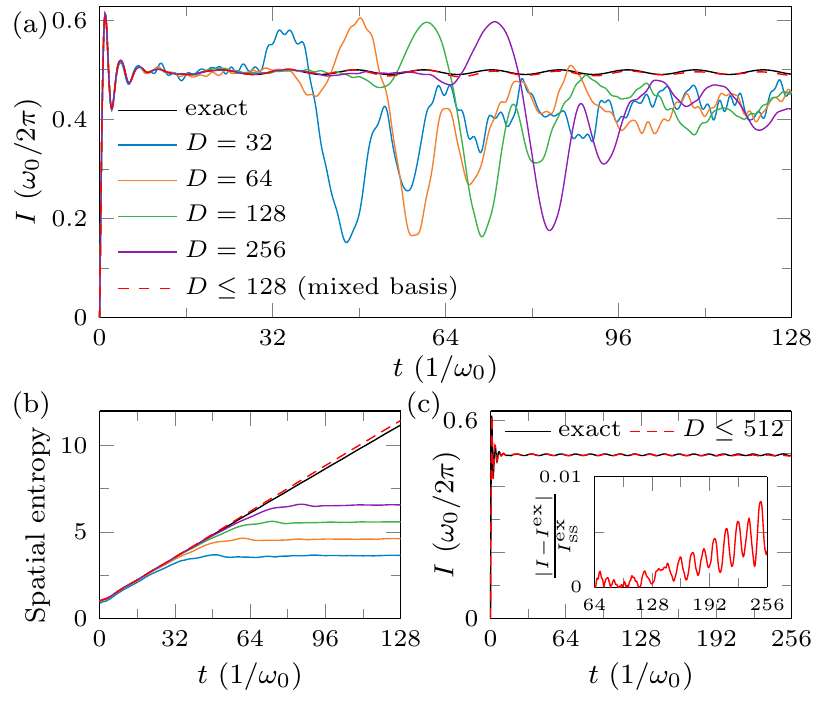}
\par\end{centering}
\caption{{\bf Failure and success.} (a) The particle current $I$ versus time for both the exact simulation via the single-particle correlation matrix (black line) and MPS simulations (colored lines). In the spatial basis, the simulation abruptly fails and successive doubling of $D$ only gives a linear increase of the achievable time scale, an exponential relationship that negates the primary advantage of MPS. In the mixed energy/spatial basis, however, a modest $D$ of 128 (necessary only around the bias window) allows for the time dynamics to be accurately captured (red, dashed line).
The Hamiltonians are $H_\s = \hbar \omega_\s \sum_\sigma n_\sigma$, $H_{\l (\r)}=\hbar\omega_0 \sum_{j\in\l (\r)} \left(b_{j}^{\dg}b_{j+1}+b_{j+1}^{\dg}b_{j}\right) + \mu_{\l (\r)} \sum_{j\in\l} b_{j}^{\dg}b_{j}$, and 
$H_\i = \hbar v \sum_{\sigma,l=1_\l,1_\r} \left(c^{\dg}_\sigma b_{l\sigma}+b_{l \sigma}^{\dg} c_\sigma \right)$ for $\s$, $\l (\r)$, and the interaction, respectively~\cite{Supplementary}.
The parameters are $v=\omega_0/\sqrt{2}$,  $\omega_S=\omega_0$, $N_\l=N_\r=256$, and  $\mu=\omega_0/2=2\mu_\l=-2\mu_\r$, where $\omega_0$ sets the frequency scale. 
For these parameters, $T(0)=1/2$, thus giving a rapid increase in spatial entanglement. The initial state has $\mu=0$ (half-filling). 
(b) Intuitively (although not precisely~\cite{verstraete_faithfully_2006,eisert_entanglement_2013}), the failure of the spatial basis is due to the conflict between a linearly growing entanglement entropy (black line, see Eq.~\eqref{eq:EntvsTime}) and the maximal amount of entanglement, $S^\star = \log_2 D$, an MPS can hold with a given bond dimension $D$ (colored lines saturating near $S^\star$). (c) Simulations of very large systems (512+1+512 sites) and long times (extensive in $N_{\l(\r)}$) are made possible by this basis, which is also reflected in the ability to capture the linear growth in spatial entanglement (red, dashed line in b). The inset shows the error in the current versus time, normalized by the exact steady-state current $I_{\mathrm{ss}}^{\rm ex}$~\cite{Supplementary}.
\label{fig:Inonint}}
\end{figure}

This growth results in the failure of one-dimensional tensor networks -- matrix product states (MPS) -- beyond a ``hard wall'': The required matrix product dimension $D$ increases {\em exponentially} with the timescale. Figure~\ref{fig:Inonint} shows this spectacular failure for the non-interacting Anderson impurity model (see the caption for its definition). This intrinsic, physically-based limitation restricts MPS to short timescales and small/moderately-sized  lattices~\cite{cazalilla_time-dependent_2002,zwolak_mixed-state_2004,gobert_real-time_2005,schneider_conductance_2006,Schmitteckert06-1,al-hassanieh_adaptive_2006,dias_da_silva_transport_2008, heidrich-meisner_real-time_2009,branschadel_conductance_2010,chien_interaction-induced_2013,gruss_energy-resolved_2018}, or linear response via an equilibrium correlation function~\cite{bohr_dmrg_2006,bohr_strong_2007}. 
Simulating time-dependent problems (artificial gauge fields, Floquet states, etc.), more complex many-body regions, or long relaxation times, requires a new approach.

The linear growth in entanglement entropy -- or its consequence, the uncontrolled growth in matrix product dimension -- is deceptive: The paradigmatic impurity model will, in the long time limit, have particles go from a state of frequency $\omega$ (on the left) to a state of the same frequency $\omega$ on the right, albeit with some characteristic spread. This entails that if one instead works with the single-particle eigenbasis of $\l$ and $\r$, ordered on a lattice as shown in Fig.~\ref{fig:Schematic}(b), the entanglement should be limited (in higher dimensions, momentum conservation can play this role). In fact, recently it was shown, in the context of dynamical mean-field theory, that for real-time single-particle correlation functions (and equilibrium) the so-called star geometry, where the energy basis for the bath is used, suppresses entanglement from logarithmic into a localized structure with smaller overall magnitude~\cite{wolf_solving_2014}. For quenches in the Anderson impurity model, it was shown that energy basis ordering naturally delineates the bad (linear entanglement growth) and good (limited entanglement) scenarios~\cite{he_entanglement_2017}.

Unlike these cases, we address simulating the bad scenario and show that it can be transformed to a scenario with logarithmic growth, and thus intermediate between the bad and good. To do so, we use a {\em mixed} energy and spatial basis, reflecting the entanglement structure in Eq.~\eqref{eq:EntProd} and incorporating the energy basis in two separate spatial regions. Figure~\ref{fig:Schematic} shows the steps leading to this mixed basis (diagonalizing the separate $\l$ and $\r$ spatial regions and then ordering them). Entanglement in this mixed basis is localized to the bias window and mostly between pairs of (iso- or nearly iso-energetic) sites, see the heat map in Fig.~\ref{fig:Schematic}(b). The strength of the couplings to the impurity, as well as many-body interactions and inelastic scattering, determine the spread of entanglement. At the same time, the low dimensionality of $\s$ and the scattering nature of the states limits the amount of  entanglement between the impurity and the reservoirs. We will comment on alternative structural representations later.

\begin{figure*}
\begin{centering}
\includegraphics[width=1\textwidth]{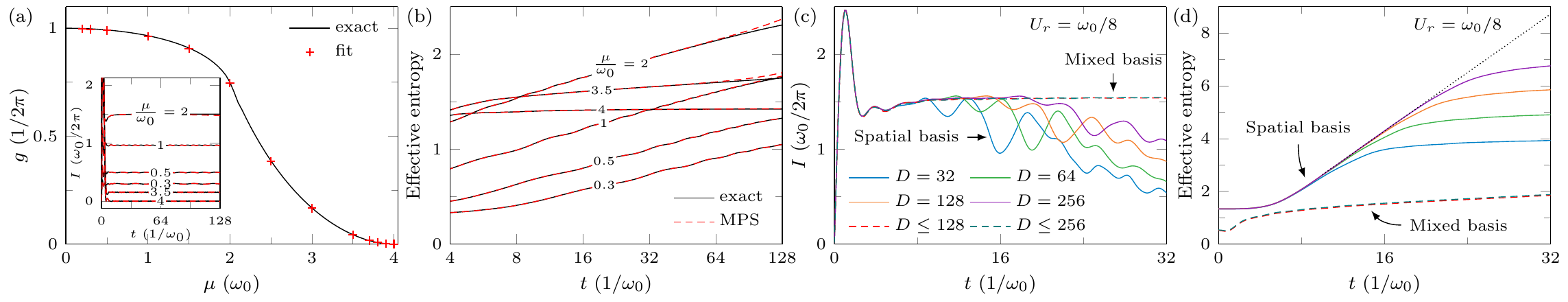}
\par\end{centering}
\caption{{\bf mixed-basis simulation.} (a,b) The conductance, $g = I_{\mathrm{ss}}/\mu$, versus bias for the mixed-basis simulation (data points) and the exact (black line) for the non-interacting model of Fig.~\ref{fig:Inonint}, as well as the entanglement growth with time for the mixed basis. We obtain the steady-state current by fitting $I(t) = I_{\mathrm{ss}} + \delta \sin(\phi t+\phi_0)$ within the window of $\omega_0 t \in [32,128]$~\cite{branschadel_conductance_2010}.
The 95 \% confidence intervals are smaller than the symbol line widths. The inset shows the current versus time for several applied biases. 
The entanglement growth is logarithmic in time for the mixed basis, unlike the linear in time growth for the spatial basis. Here, $N_\l=N_\r=256$ and $D = 256$ except for $\mu/\omega_0=1$ and 2, for which $D = 512$. (c,d) The current versus time for interacting $\l$ and $\r$ reservoirs that have the additional contribution $\hbar U_r \sum_{j\in \l \r} (n_j-1/2) (n_{j+1}-1/2)$ to the  Hamiltonian, as well as the effective entanglement entropy for the spatial and mixed bases (black, dotted line extrapolates the linear region of the entropy for the largest $D$). The MPO of the Hamiltonian increases when $U_r\neq 0$, but the mixed-basis simulation enables stable evolution that is not possible in the spatial basis. The advantage of the mixed basis becomes substantial as $\l\r$ entanglement increases with time or with the transport parameters (e.g., with the bias). For this interacting model, the results are for $\mu/\omega_0 = 2$ and $N_\l=N_\r=64$. \label{fig:Inewmethod}}
\end{figure*}

We note here that various approaches can perform computations with matrix product states, such as the density matrix renormalization group (DMRG)~\cite{white_density_1992}, the time-dependent variational principle (TDVP)~\cite{haegeman_unifying_2016}, or Krylov-based methods \cite{zaletel_time-evolving_2015}. Unlike the schemes based on the Trotter decomposition of the Hamiltonian into local gates, they allow treating any Hamiltonian represented as a matrix product operator (MPO). Thus, we use DMRG to find the initial ground state in the preferred basis and TDVP for the subsequent time evolution. Since we work with the MPO of the Hamiltonian, its dimension is important since the formal scaling for time evolution in, e.g., 1D is $\mathcal{O} (D^3 M d + D^2 M^2 d^2)$, where $M$ is the MPO bond dimension and $d$ is the local Hilbert space dimension. When the reservoirs are non-interacting, the MPO of the mixed basis has a small, fixed $M$ for both the initial state and time evolution regardless of bias~\cite{Supplementary}. 

Since there is overhead associated with the presence of long-range interactions, we also work under a guiding principle that both the Hamiltonian MPO and the state MPS should have limited $D$. The mixed basis, in contrast to the spatial basis (exponentially large MPS $D$) and the global single particle basis (extensive MPO $D$ when interactions are present), respects this principle in addition to capturing the natural structure of impurity transport~\cite{Supplementary}.
Optimality questions aside, it permits an advantageous extension to open systems~\cite{2019inprep} where a bias is maintained by external contacts to $\l$ and $\r$ eigenstates separately~\cite{gruss_landauers_2016,gruss_communication:_2017,elenewski_communication:_2017}, as well as a suitable structure for fine--graining the reservoirs~\cite{Zwolak08-2}. 

Figure~\ref{fig:Inonint}(a) shows the result of employing this mixed-basis MPS. An inhomogeneous and modest $D \le 128$ already gives excellent results out to a time extensive in the system size (a time equal to the reservoir length divided by the Fermi velocity, $2 \omega_0$, at which the current ``front'' hits the open boundary and travels backwards toward the impurity~\cite{chien_landauer_2014,chien_bosonic_2012}). This mixed basis captures the linear growth in spatial entanglement entropy, Fig.~\ref{fig:Inonint}(b), and allows for very large systems, Fig. \ref{fig:Inonint}(c). 

As a consequence of naturally representing the entanglement structure, the majority of the lattice in the mixed basis has little entanglement across bipartite cuts with correlations predominantly between modes in the bias window (see Fig.~\ref{fig:Schematic}(b)). Thus, the {\em computational speedup} is not just a consequence of a reduced $D$, but also an inhomogeneous $D$. As a point of comparison, the mixed-basis simulations in Fig.~\ref{fig:Inonint}(a) took only 15 hours, whereas the spatial-basis simulation with the same $D=128$ took 44 hours, both on the same single core computer. 
While implementation choices affect this comparison, the empirical scaling follows from Fig.~\ref{fig:Inonint}: To bring the spatial-basis simulation out to $t=128 \, \omega_0$ requires five more doublings of $D$ just to move the breaking point (forgetting about overall error). The dominant $D^3 M d$ contribution to the computational cost then indicates an approximate computational time of $(2^5)^3\cdot44$ hours, or {\em 165 years}.

\begin{figure*}
\begin{centering}
\includegraphics[width=1\textwidth]{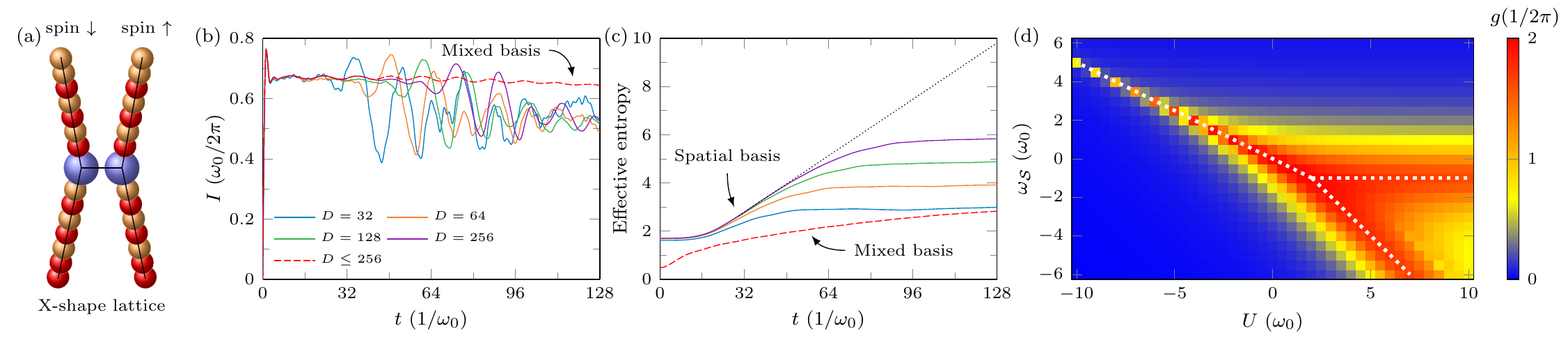}
\par\end{centering}
\caption{{\bf Many-body transport in a mixed, multi-channel basis.} (a) The ``X-lattice'' we employ when multiple channels are present. (b) Current versus time for spatial- and mixed-basis simulations, showing 
abrupt failure of the spatial basis. (c) Effective entanglement versus time, which accounts also for the inhomogeneity of $D$~\cite{Supplementary}. The spatial basis shows a linear growth in spatial entanglement (black, dotted line extrapolates the linear region of the entropy for the largest $D$), similarly to its non--interacting counterpart. The mixed basis, however, has only moderate growth in entanglement and essentially does not change as $D$ increases beyond the value shown. (d) Conductance diagram for the Anderson impurity model versus the system's on-site frequency $\omega_\s$ and interaction strength $U$. White dashed lines indicate high conductance states, see the main text for discussion.
The steady state is found in the same way as Fig.~\ref{fig:Inewmethod}(a) but using the window $\omega_0 t \in [32,64]$ for $N_\l=N_\r=128$. The parameters are $v = \omega_0/\sqrt{2}$ and $\mu = \omega_0/5$. Note that the broken $U<0$ line is due to the resolution of the figure.   For (b) and (c), $N_\l=N_\r=256$, $U=5 \, \omega_0$, $\omega_S = 0$, and $\mu = \omega_0/2$.
\label{fig:anderson1}}
\end{figure*}

Even for a large bias, the mixed-basis MPS still out performs the spatial basis, a direct consequence of the more local nature of entanglement in energy. Figure~\ref{fig:Inewmethod}(a) shows the conductance and  current versus time traces for $\mu = 0$ to $4 \, \omega_0$ (after which the $\l$ and $\r$ bands go out of alignment and there is no steady state current and entanglement saturates). In all cases, the mixed basis yields accurate results. Figure~\ref{fig:Inewmethod}(b) shows that the effective entanglement entropy~\cite{Supplementary} grows logarithmically opposed to linear. The advantages of the mixed basis also extend  to interacting reservoir models, where Fig.~\ref{fig:Inewmethod}(c,d) show the current versus time and effective entanglement entropy for the same $\l\s\r$ system but with interactions in $\l\r$. The Hamiltonian's MPO dimension grows with the system size in this case, and in some parameter regimes we expect a spatial basis may be more suitable (such as in a localized regime), but the mixed basis enables stable time evolution. Further work will be necessary to assess the efficiency gains across different parameter regimes, since many-body interactions in $\l$ and $\r$ modify both the MPO and entanglement structure of the problem.

The above can be straightforwardly applied to higher dimensional non-interacting reservoirs (since only their energy/momentum bases matter) and to interacting spinless fermions, including larger dimensional systems $\s$. For many-body systems of typical interest, though, one has to have spin, which requires simulating multiple channels. The Anderson impurity problem~\cite{anderson_localized_1961} with electron-electron interaction $U n_\uparrow n_\downarrow$ at the impurity (where $n_{\uparrow(\downarrow)}$ is $\s$'s spin up (down) number operator), and its extension to larger $\s$~\cite{bauer_microscopic_2013,iqbal_odd_2013}, is the paradigmatic example. 

In the non-interacting limit the two spin channels are fully independent. In the presence of interactions, $U\neq0$, the interchannel entanglement originates from the many-body contact at $\mathcal{S}$, which is in addition to the intrachannel entanglement around the bias window~\cite{Supplementary}. This suggests an X-shape MPS (i.e., a tree tensor network \cite{Vidal_tree_2006,Murg_tree_2010}) depicted in Fig.~\ref{fig:anderson1}(a) as a natural ansatz (a structure also supported by results of other recent works~\cite{bauernfeind_fork_2017}). Tree tensor networks, similarly to a one-dimensional MPS, possess a normal form. As such, the TDVP integration scheme of Ref.~\cite{haegeman_unifying_2016} directly extends to such tensor network geometries~\cite{Supplementary}. 

Figure~\ref{fig:anderson1}(b) shows the X-lattice simulated in both the spatial and mixed bases. Just as with the non-interacting case, the spatial basis abruptly fails and increasing $D$ only gives logarithmic increase in the achievable simulation time. The mixed basis, though, enables the simulation to go out to a time extensive in the reservoir size. When $D$ is too small, it will lose accuracy, but it does not abruptly fail. This is reflected in the limited growth in entanglement, Fig.~\ref{fig:anderson1}(c), which behaves similarly to the non-interacting case.

Figure~\ref{fig:anderson1}(d) shows the conductance diagram of the paradigmatic Anderson impurity problem. For negative $U$, $\s$ is approximately half occupied in each channel, giving a $U/2$ contribution to $\s$'s on-site energy in the other channel. This results in a single high conductance state when the level energy is pushed into the bias window at $\omega_\s \approx -U/2$. As $\omega_\s$ becomes negative, the conductance peak bifurcates into two particle-hole dual, correlated high-conductance states. For one, there is a correlated state between one channel being occupied and current flowing in the other channel, giving $\omega_\s \approx -U$ to effectively push the current-carrying channel state into the bias window. The other is a correlated state between one channel being empty and current flowing in the other channel. This occurs at $\omega_\s \approx -\omega_0$ instead of $\omega_\s=0$ due to residual many-body correlations increasing the energy (a residual also present in the $\omega_\s \approx -U$ state). The accurate calculation of the whole conductance diagram enables the identification of these features.

Finally, we comment on the computational speedup. The spatial basis requires exponentially large $D$ in the total simulation time $T$, already alluded to above: Each doubling of $D$ increments the breaking point by $\Delta t$ (independent of the value of $D$), giving $\Delta t \ln D \approx T$ or $D\approx e^{T/\Delta t}$. For the mixed-basis simulation of the single channel model, we examine the error versus $D$ for several simultaneous multiples of the reservoir size and time~\cite{Supplementary}. The simultaneously, e.g., doubling of size and time is the method by which one achieves the long-time limit. The error decay suggests that doubling of the simulation time (and size) requires increasing $D$ to $\alpha D$, where $\alpha$ is bounded, to keep the overall error fixed. Thus, the computational cost is brought from $e^{3T/\Delta t}$ to $T^p$, where $p \approx 3$~\cite{Supplementary}. We note, however, that for fermions with spin, the X-lattice configuration requires also evolution between the two channels. The entropy across this bond is the same in the spatial and mixed basis, and can itself increase linearly in time. For the range of parameters here, it is still quite small. In principle, this will dominate the scaling for long times. However, one will still get an exponential improvement in the {\em prefactor} of this contribution, since that prefactor depends in the intrachannel entanglement and thus is suppressed when going to the mixed basis. Other structures besides the X-lattice may improve this further.

The above general considerations demonstrate that difficult computational problems can be broached so long as the natural entanglement structure is recognized -- here, by changing the canonical basis. This enables the accurate simulation of quantum transport that underlies many applications, from quantum dot platforms for computing to molecular and nanoscale electronic devices, and fundamental studies with cold-atom emulators. The long times achievable will be conducive to simulating transport through systems undergoing time-dependent driving to generate artificial gauge fields or Floquet states. Combining the approach here with other recent methods~\cite{Dora_master_2015,schwarz_nonequilibrium_2018,Fugger_kondo_magnetic_2018} will push further the limits of simulation, as will  developing algorithms to locally optimize the canonical basis~\cite{krumnow_fermionic_2016}. As such, our results open new avenues to study the behavior and simulation of non-equilibrium many-body systems, from fermionic impurities to bosonic baths to the inherent structure of tensor networks. 

\begin{acknowledgments}
We thank Y. Dubi, M. Ochoa, J. A. Liddle, and J. Elenewski for comments. M.M.R thanks F. Verstraete for inspiring discussions and acknowledges support by National Science Center, Poland under Projects No. 2016/23/D/ST3/00384.
After completing this work, we became aware of Ref.~\onlinecite{Jens2019}, which takes a different technical approach, but is close in mindset and motivation.
\end{acknowledgments}
%

\vspace{0.5cm}
\onecolumngrid
\renewcommand{\theequation}{S\arabic{equation}}
\setcounter{equation}{0}
\renewcommand{\thefigure}{S\arabic{figure}}
\setcounter{figure}{0}
\section{Supplementary Material}
Quantum transport through an impurity or interface is typically approached via the Hamiltonian~\cite{jauho_time-dependent_1994,caroli_direct_1971} 
\begin{equation}
H = H_{\s} + H_{\i} + H_{\l} + H_{\r}, \label{eq:H}
\end{equation}
where $H_\s$ is the many-body Hamiltonian of the impurity region -- the system $\s$ -- which may include electron-electron interactions, electron-phonon/vibrational coupling, etc. The remaining terms are the coupling of the system and reservoirs, and the isolated reservoir Hamiltonians. Reflecting the non-interacting nature of the Fermi sea and recognizing that the partitioning into $\s$ and $\l\r$ can be done so that the relevant, spatially localized interaction region is in $\s$, it is standard~\cite{jauho_time-dependent_1994,caroli_direct_1971} to take these other Hamiltonians as quadratic forms
\begin{equation}
H_{\i}=\sum_{i\in\s,k\in\l\r}\hbar v_{ik}\left(c_{i}^{\dg}a_k+a_k^{\dg}c_{i}\right) \label{eq:HI}
\end{equation}
and 
\begin{equation}
H_{\l (\r)}=\sum_{k\in\l (\r)}\hbar\omega_{k}a_{k}^{\dg}a_{k}, \label{eq:HL}
\end{equation}
where $c_{i}^{\dg}$ ($c_{i}$) and $a_k^{\dg}$ ($a_k$) are the creation (annihilation) operators in $\s$ and $\l\r$, respectively, and $v_{ik}$ is the coupling for modes $i \in \s$ and $k \in \l\r$. Spin (when present) is implicit in the labels. This ``impurity'' Hamiltonian is the same general structure as that addressed with the Numerical Renormalization Group (of course, one can modify Eq.~\eqref{eq:HI} to have some other $\s$ operator but retaining the linearity in $\l\r$ operators). There, a logarithmic discretization of the energy basis in the reservoir(s) gives a finite number of modes with a more fine structure at low energy. After a transformation of this Hamiltonian with a finite number of degrees of freedom to a one-dimensional lattice (a Wilson chain), an iterative diagonalization process yields the low energy states~\cite{Wilson75-1,bulla_numerical_2008}.

Matrix product state simulations require a (quasi-) 1D lattice. Prior simulations thus considered either explicitly a lattice in one spatial dimension~\cite{cazalilla_time-dependent_2002,zwolak_mixed-state_2004,schneider_conductance_2006,Schmitteckert06-1,al-hassanieh_adaptive_2006,dias_da_silva_transport_2008,heidrich-meisner_real-time_2009,branschadel_conductance_2010,chien_interaction-induced_2013,gruss_energy-resolved_2018} or some other real-space-like construction (e.g., one spatial dimension with energetically tapered boundaries~\cite{dias_da_silva_transport_2008,branschadel_conductance_2010}). Essentially, this amounts to considering the reservoir Hamiltonian
\begin{equation}
H_{\l}=\hbar\omega_0 \sum_{j\in\l} \left(b_{j}^{\dg}b_{j+1}+b_{j+1}^{\dg}b_{j}\right) + \hbar \mu_{\l} \sum_{j\in\l} b_{j}^{\dg}b_{j} \label{eq:HL1D}
\end{equation}
and similarly for $H_\r$. For simplicity, we take the hopping and chemical potential to be uniform within each reservoir, and take the same $\omega_0$ for both $\l$ and $\r$. The $b_{j}$ ($b_{j}^{\dg}$) are the creation (annihilation) operators in $\l\r$ at the real-space site $j$.

When the system is a single Anderson impurity with equal coupling $v$ to both reservoirs, the remaining Hamiltonians are 
\begin{equation}
H_\s = \hbar \omega_\s \sum_\sigma n_\sigma + \hbar U n_\uparrow n_\downarrow \label{eq:HS}
\end{equation}
and
\begin{equation}
H_\i = \hbar v \sum_{\sigma,l=1_\l,1_\r} \left(c^{\dg}_\sigma b_{l\sigma}+b_{l \sigma}^{\dg} c_\sigma \right),
\end{equation}
where all indices now explicitly include spin $\sigma$ [Eqs.~\eqref{eq:H}--\eqref{eq:HL1D} have spin implicit], $n_\sigma=c_\sigma^\dg c_\sigma$ is the number operator on the system site with spin $\sigma$, and $l=1_{\l (\r)}$ is the site in $\l$ ($\r$) that contacts $\s$. For the specific simulations in this work, we will use this model and vary $\omega_S$, $v$, and $U$. However, we will work in the single-particle eigenbasis of each of these reservoirs separately. Thus, {\em the spatial nature of this lattice will be inconsequential to the general considerations in our work} (it only will change the band structure and the dispersion of the coupling). The computational approach can thus handle non-interacting reservoirs in any dimension, 1D, 2D, 3D, etc., and with long-range hopping. 

The canonical transformation that defines the eigenbasis for the 1D reservoir model is $a_k=\sum_{j\in\l}\u^\dg_{kj} b_j$ with  $\u^\dg_{kj}=\sqrt{2/\left(N+1\right)} \sin\left(jk\pi/\left(N+1\right)\right)$ and $k=1,\ldots,N$ for an $N$ site reservoir, yielding  
\begin{equation}
\omega_{k}= 2\omega_0\cos\left(k\pi/\left(N+1\right)\right)+\mu_{\l},
\label{eq:omega_k}
\end{equation} 
and $v_k=v\,\u^\dg_{k1}$ in Eq.~\eqref{eq:HI} (and similarly for $\r$). The simulation technique will work in the more general setting where $\s$ is an arbitrary interacting impurity with many electronic sites and vibrational modes, although the computational cost will depend on the Hilbert space dimension and structure of $\s$.

To drive a current, we consider the $\l \s \r$ system initially in contact and in its ground state at zero temperature. At time $t=0$, a bias $\mu_\l=-\mu_r=\mu/2$ turns on, generating a current. An alternative case is to have $H_\i=0$ and $\mu_\l=-\mu_r=\mu/2$ initially, then turn on $H_\i$ and off $\mu$. This starts the system with a density imbalance that drives the current when the applied chemical potential no longer sustains the imbalance. A third case is to have $H_\i$ on initially and also the chemical potential drop, letting the latter go to drive the current. These lead to different time dynamics and initial entanglement, but they yield the same steady state and asymptotic entanglement growth~\cite{chien_landauer_2014}. The current from left to right is
\begin{equation}
I(t)=-\avg{dn_{\l}/dt}=-2\sum_{\sigma,k\in\l}v_{k}\Im\avg{a^\dg_{k} c_\sigma}, \label{eq:Current}
\end{equation}
where $n_\l$ is the total particle number on the left reservoir, $\Im$ is the imaginary component, and again spin is implicit in the label $k$. In all simulations, the overall filling is determined by the initial state. It is set at (almost) half-filling with $N$ electrons for a system of $N+1+N$ modes (per spin channel). 

For any finite system, one can directly simulate the dynamics of non-interacting electrons by evolving the correlation matrix~\cite{elenewski_communication:_2017}
\begin{equation}
\c_{mn} = \tr \, [ d^\dg_m d_n \, \rho ] ,
\end{equation}
where $\rho$ is the full density matrix and $d^\dg_m$ ($d_n$) are the creation (annihilation) operators at mode $m,n \in \l \s \r$. Defining the single-particle Hamiltonian $\bar{H}$ through
\begin{equation}
H=\sum_{m,n \in \l \s \r} \bar{H}_{n m} d^\dg_m d_n 
\end{equation}
and using that $\tr \left( d^\dg_m d_n [H,\rho] \right) =  [\bar{H},\c]_{mn}$, the evolution of the correlation matrix is 
\begin{equation}
\label{eq:C}
\dot{\c} = -\im [\bar{H},\c]/\hbar .
\end{equation}
This equation can be evolved directly. The dynamics can also be simulated by diagonalizing the ``small'' dimensional $\bar{H}$ and then transforming the correlation matrix into its eigenbasis~\cite{chien_landauer_2014}.

We first consider the fully non-interacting model [$U=0$ in Eq.~\eqref{eq:HS}], dropping also the spin since there is no interaction between spin channels (we will multiply the current by an additional factor of two to account for spin, which is not the factor of two already appearing in Eq.~\eqref{eq:Current}). We then consider the case with interactions. 

Figure~\ref{fig:Inonint} of the main text shows a matrix product state (MPS) and an exact (via the correlation matrices) simulation of transport in the non-interacting model, Eq.~\eqref{eq:HS} with $U=0$, using the spatial basis, as has been done in prior work. The steady-state particle current, $I_{\mathrm{ss}}$, is given by Landauer's formula (regardless of the protocol for driving the current),
\begin{eqnarray}
I_{\mathrm{ss}}^{\rm ex}&=&\frac{1}{\pi}\int_{-\infty}^{\infty}d\omega(f_{\l}(\omega)-f_{\r}(\omega))T(\omega) = g^{\rm ex} \mu \label{eq:Landauer} \\
&\approx& \frac{1}{\pi} T(0) \mu,
\end{eqnarray}
where we explicitly include the factor of two out front (cancelling a factor of 1/2) to account for both spin channels and the second line is in linear response. This equation also defines the exact value of conductance $g^{\rm ex}$ for non--interacting case. The $f_{\l (\r)}$ are the Fermi-Dirac distributions in the left (right) reservoir. The transmission function is 
\begin{equation}
T(\omega)=\Gamma_{L}\Gamma_{R}|G(\omega)|^{2},\label{eq:T}
\end{equation}
with the retarded Green's function of the impurity $G(\omega)=1/[\omega-\omega_S-\Sigma_{\s\l}-\Sigma_{\s\r}]$, the spectral function of the couplings $\Gamma_{\l (\r)}=-2\Im \Sigma_{\s\l (\s\r)}$, the self-energies $\Sigma_{\s\l (\s\r)}=v^{2}G_{\l (\r)}(\omega)$, and the reservoir Green's functions $G_{\l (\r)}(\omega)=1/[\omega-\mu_{\l (\r)}-\Sigma_{\l (\r)}(\omega)]$, and self-energies $\Sigma_{\l (\r)}=(1/2)\left[\omega-\mu_{\l (\r)} - \imath \sqrt{4\omega_0^{2}-(\omega-\mu_{\l (\r)})^{2}}\right]$.

The response of the total system to the driving force results in a rapid rise of the current from zero as particles flow from one reservoir to the other, going into oscillations (due to the presence of Gibbs phenomenon) that decay as the current goes into a quasi-steady state~\cite{zwolak_communication:_2018}. With a large matrix product dimension, the current from the MPS simulation will match the exact solution reasonably well until it abruptly fails for the spatial basis. The origin of the failure is the scattering nature of the problem: Particles come in from, e.g., the left, scattering off the interface at the impurity, generating entanglement between the two reservoirs in the process.

\begin{figure}[t]
\begin{centering}
\includegraphics[width=0.49\columnwidth]{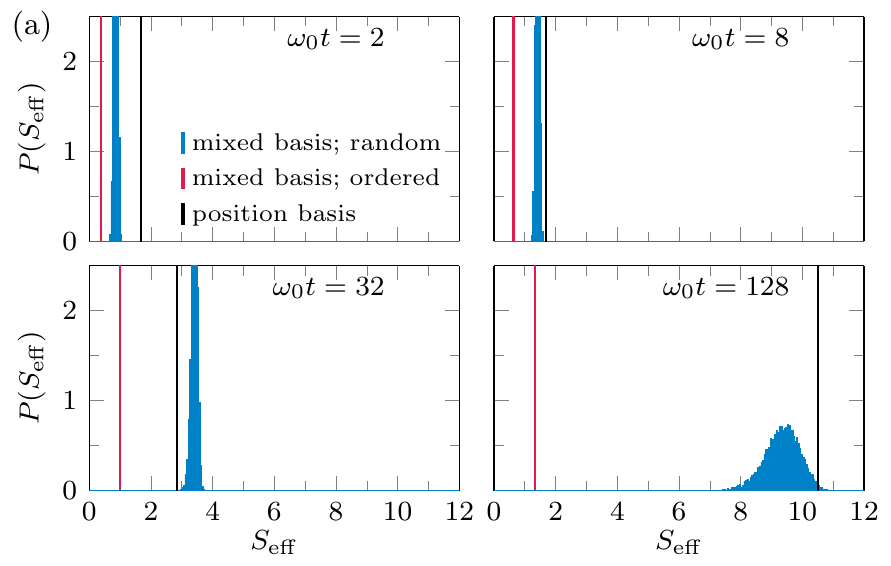}
\includegraphics[width=0.49\columnwidth]{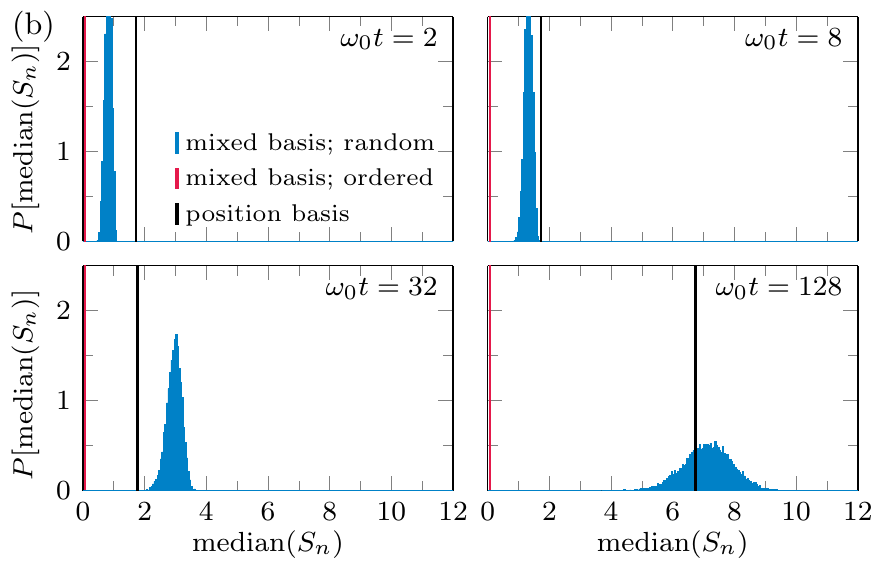}
\end{centering}
\vspace{-5pt}
\caption{{\bf Effective entropy and mode ordering in the mixed basis.} For the setup in Fig.~\ref{fig:Inonint} of the main text with $N_\l=N_\r=256$, we compare the entanglement entropy during different stages of the evolution in the position basis, mixed basis with modes ordered according to energy, and the mixed basis with random ordering. For the latter, we show a histogram of $10^4$ orderings. The plots show the probability for the {(a)} effective entropy $S_{\rm eff}$ (blue histogram), see Eq.~\eqref{eq:Seff}, and {(b)} median entropy of the possible bipartite divisions of the 1D lattice. The expected entropy for random orderings in the energy basis is comparable to the position basis, being both large and growing linearly in time (in the position basis, there is an initial, residual entropy due to the fact that the initial state is a critical state with entanglement logarithmic in the system size). For all but the shortest times, both are significantly larger than the energy-ordered mixed basis. The median entropy shows that the result for the effective entropy is not due to a small subset of large entropy cuts. Moreover, the median shows that the energy-ordered mixed basis has basically no entanglement throughout the lattice. These results clearly indicate that the right ordering is crucial for properly capturing the entanglement structure of the system during all stages of the evolution.\label{fig:S2}}
\end{figure}

\begin{figure}[t]
\begin{centering}
\includegraphics[width=0.95\columnwidth]{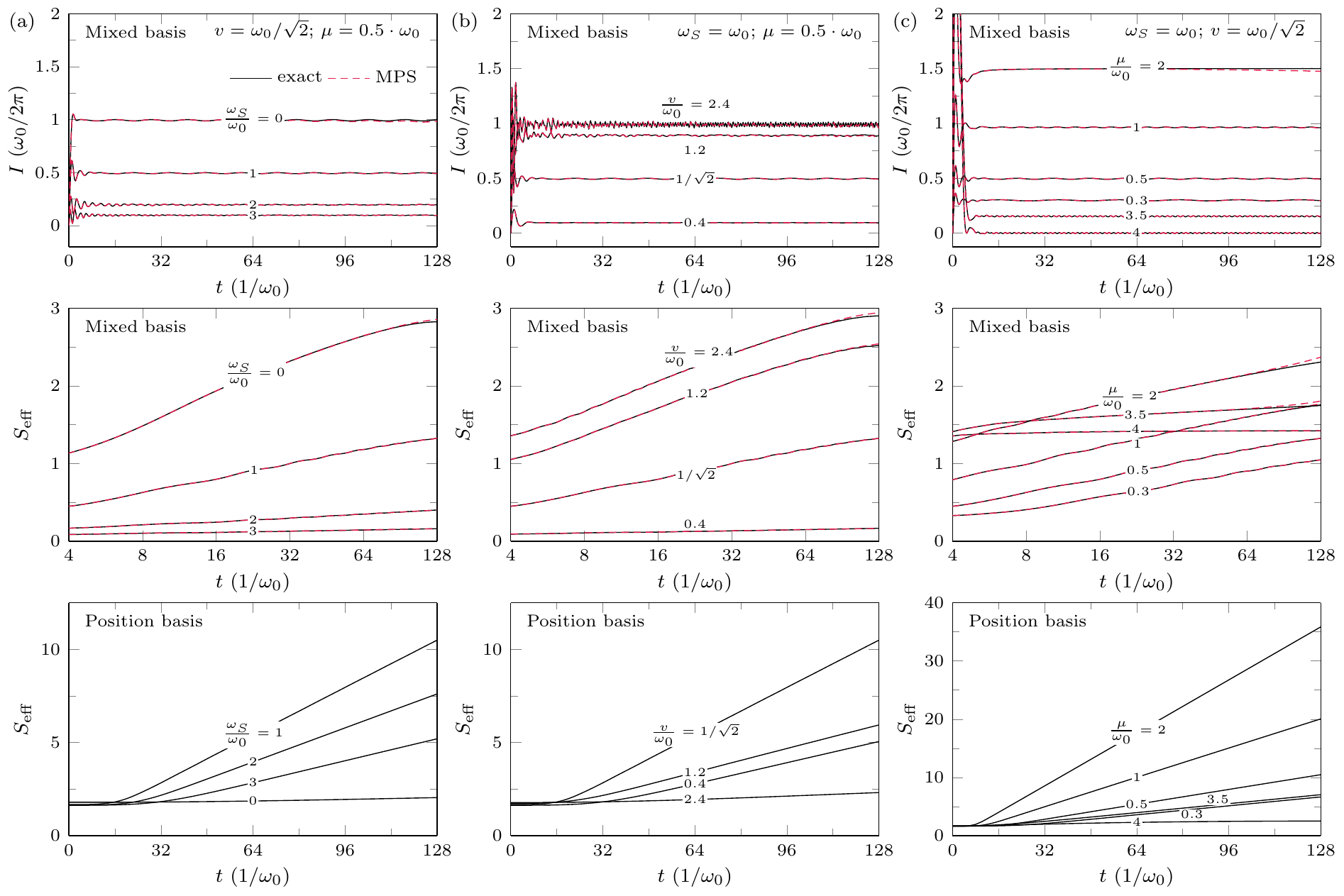}
\end{centering}
\vspace{-5pt}
\caption{{\bf Current and entanglement for the mixed basis at $U=0$.} We show complementary results to the ones in Fig.~\ref{fig:Inewmethod} of the main text. Here, $N_\l = N_\r = 256$ and $D\le 256$ (apart from $\mu/\omega_0 = 1$ and $2$ in column (c) -- where, for completeness, we repeat part of Fig.~\ref{fig:Inewmethod} of the main text --  see the caption of that figure). We also show the effective entropy, Eq.~\eqref{eq:Seff}, in the position and mixed bases for different slices of (a) $\omega_\s$,  (b) $v$, (c) $\mu$. There are special situations -- namely, resonant transport -- where no entanglement is generated in the position basis as there is no scattering. These situations, though, are of measure zero in parameter space. \label{fig:S3}}
\end{figure}

Figure~\ref{fig:S2} shows the maximum bipartite entanglement entropy across the lattice for position basis and different orderings in mixed basis.
The ``natural'' ordering, with reservoir modes paired with their closest frequency mode on the opposing side and with the system placed around the Fermi level, has the smallest entanglement. We choose to consider canonical transformations that are only permutations to ensure that the matrix product operator that defines the evolution is low dimensional (see below), and, in particular, does not grow with time. 
We note that for the particular case with a fully non--interacting model, the global single-particle eigenbasis of the Hamiltonian for the time dynamics has zero entanglement growth during evolution. However, the MPO dimension for the initial Hamiltonian grows linearly with the total lattice size (and rotating the interacting component further adds to the MPO dimension). Having interacting systems in mind, including ones with larger interacting regions $\s$, we limit ourselves to the most natural mixed basis, which allows both for a simple MPO and -- via a proper ordering -- representation of the entanglement structure, in correspondence with the guiding principle mentioned in the main text. 
In Fig.~\ref{fig:S3}, we show the results, including the entanglement entropy, of the mixed basis simulations for other values of $\omega_S$, $v$, $\mu$, and $U=0$ (complementary to Fig.~\ref{fig:Inewmethod} of the main text).

\begin{figure}[t]
\begin{centering}
\includegraphics[width=0.9\textwidth]{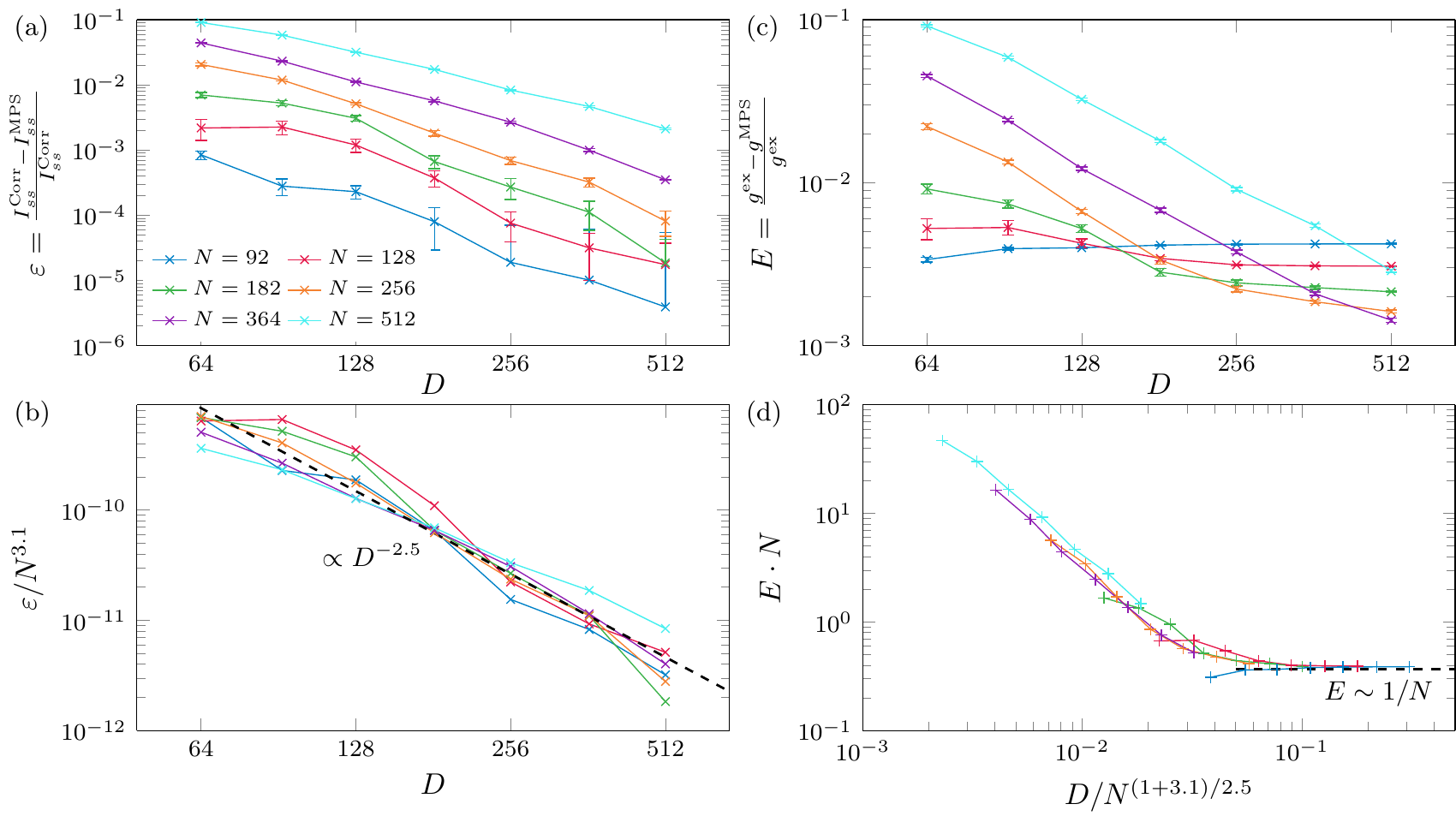}
\par\end{centering}
\vspace{-5pt}
\caption{{\bf Error in the steady state current and conductance}. {(a)} We consider the setup in Fig.~\ref{fig:Inonint} of the main text. For different system sizes $N_\l=N_\r=N$, we obtain the steady-state current by fitting $I(t) = I_{\mathrm{ss}} + \delta \sin(\phi t+\phi_0)$ within the  window of $\omega_0 t \in [N/4,N/2]$. 
This simultaneously increases both lattice length and the timescale, which is the normal process by which the long-time limit is taken. The error bars indicate 95 \% confidence intervals of the fits. We compare the extracted steady-state current, $I^{\mathrm{MPS}}_{ss}$, of simulations for different values of MPS bond dimension $D$ with the exact value extracted from, $I^{\mathrm{Corr}}_{ss}$, from evolving the single-particle correlation matrix, with the measure of error taken as the relative difference, $\varepsilon$, with the exact value. This comparison shows that to take the long-time limit (e.g., successive doubling of both time and space) at a fixed simulation error, the necessary $D$ only increases by approximately a constant factor. The error is not expected to decrease by simple power laws and the waviness in the plots is due to features in the spectrum of Schmidt coefficients. Other factors that influence the error (timestep, Schmidt tolerance, etc.) -- ones that might have some complicated interplay with the matrix product dimension -- are taken to be sufficiently stringent to give only a secondary influence in the error versus $D$.
Only when taking the asymptotic limit and coarse--graining the error (smoothing over some range of $D$'s) might a simple decay emerge. Nevertheless, in {(b)} we attempt power law fit to the above data, obtaining the leading behavior 
$\epsilon \sim e^c L^a / D^b$, with $a=3.1\pm 0.3$, $b=2.5 \pm0.2$ and $c=-10.5\pm1.5$.
{(c)} The relative error in the conductance $E$ compared to that found from the exact calculation in the infinite system/time limit (see Eq.~\eqref{eq:Landauer}, with $g^{\mathrm ex}=I^{\rm ex}_{\rm ss}/\mu$). Unlike in (a), this error has a minimum value for a fixed finite size and time due to an offset from the infinite system/time result~\cite{branschadel_conductance_2010,zwolak_communication:_2018}. That is, this error (for an exact simulation of a finite system) vanishes as $1/N$. Increasing $D \to \infty$ will therefore not remove this error. Rather, the error will decay initially as $D$ increases, but will level off when reaching this asymptotic value (potentially, as seen for $L=92$, increasing to this limit). Increasing the accuracy of the calculation requires a simultaneous increase of the length- and time-scales, as well as the matrix product dimension. {(d)} We use the above to rescale the overall error, obtaining reasonable collapse for different $D$ and $N$. The optimal combination of the latter to reach given quality of $E$ corresponds to the crossover point. \label{fig:S1}
}
\end{figure}

{\em Many-body simulation.} 
Various approaches can perform computations with matrix product states, such as the density matrix renormalization group (DMRG)~\cite{white_density_1992} for obtaining the ground state, and the time-evolving block decimation algorithm (TEBD)~\cite{vidal_efficient_2003,vidal_efficient_2004}, Krylov- and expansion-based methods~\cite{schmitteckert_nonequilibrium_2004,zaletel_time-evolving_2015} and the time-dependent variational principle (TDVP)~\cite{haegeman_time-dependent_2011,haegeman_unifying_2016} for simulating time evolution.
In order to simulate the time evolution, we employ TDVP for matrix product states (MPS)~\cite{haegeman_time-dependent_2011,haegeman_unifying_2016}, where for simulating the effects of local gates we apply the Krylov-based method of Ref.~\onlinecite{Niesen_Krylov} that is adaptive both in the timestep and in Krylov-subspace dimension.
TDVP provides a means to tackle a broad class of Hamiltonians represented as a matrix product operator (MPO). For instance, the Hamiltonian in Eq.~\eqref{eq:H} -- limited to a single spin channel ($U = 0$) in the mixed energy-spatial basis, Eqs.~\eqref{eq:HL} and \eqref{eq:HS}, and a single system site interacting with the reservoirs -- can be expressed as an MPO with a small bond dimension, $M=4$ (for the spatial Hamiltonian in Eq.~\eqref{eq:HL1D} and without interactions in $\s$, the bond dimension is also 4). The latter is independent on the particular ordering of the energy modes. Each additional system site interacting with the reservoir would increase this bond dimension by 2. For a single channel and a single site in $\s$, the exact form for the mixed basis is 
\begin{equation}
H = \prod_{\omega_k<0} W^{k} \cdot W^{\mathcal{S}} \cdot \prod_{\omega_k>0} W^{k},
\label{eq:MPOH}
\end{equation}
where the sites are ordered according to $\omega_k$ form Eq.~\eqref{eq:omega_k} (jointly for $\l$ and $\r$).  The initial state is the ground state of the Hamiltonian with $\mu_L = \mu_R = 0$, which has an MPO of the same form (with sites ordered using the nonzero value of $\mu$ for $t>0$ so that exactly the same basis is used for the initial state and the subsequent time evolution). Finally, the MPO matrices read
$$ W^{\omega_k<0} = \begin{pmatrix}  
\mathds{1} &  v_{ik}  a_k  &  v_{ik}  a^\dagger_k &  \omega_k a_k^\dagger a_k   \\
   & \mathds{1}&   & \\
   &   &\mathds{1} & \\
   &   &   & \mathds{1} \\
\end{pmatrix},
$$
$$W^{\mathcal{S}} = \begin{pmatrix}  
\mathds{1} & - c_i  &  c_i^\dagger & \omega_\mathcal{S} n_i  \\
   & &   & - c_i^\dagger \\
   &   &   &c_i  \\
   &   &   & \mathds{1} \\
\end{pmatrix}, 
$$
and
$$
W^{\omega_k > 0} = \begin{pmatrix}  
\mathds{1} &  &  & \omega_k a_k^\dagger a_k  \\
   & \mathds{1}&   &  v_{ik} a_k^\dagger \\
   &   &   \mathds{1}& v_{ik}  a_k  \\
   &   &   & \mathds{1} \\
\end{pmatrix}. 
$$
The terms equal to zero have been left blank to show the sparsity of the these matrices. Finally, the first and the last matrix in Eq.~\eqref{eq:MPOH} (i.e., the smallest and largest energy modes) are limited only to the first row and column, respectively.  For such a setup, one can also employ the Jordan-Wigner transformation to the pseudo-spin operators. The fermionic nature of the problem is then reflected by, among other things, $\sigma^z$ operators replacing the identities $\mathds{1}$ that connect separate creation and annihilation operators.
For the case of interacting reservoirs in Fig.~\ref{fig:Inewmethod}(c,d) of the main text, the additional contribution to the Hamiltonian reads $\hbar U_r \sum_{j\in \l \r} (n_j-1/2) (n_{j+1}-1/2)$. Its MPO in the mixed basis is generated starting from the simple MPOs representing elementary operators. The full Hamiltonian is obtained by subsequent multiplication and addition (as well as bond dimension compression of the  resulting MPOs) of the elemental components, using the standard calculus of matrix product states, see e.g., Ref.~\cite{schollwock_density-matrix_2011}. Its bond dimension grows approximately as $\sim 24 N$.

\begin{figure}[t]
\begin{centering}
\includegraphics[width=0.40\columnwidth]{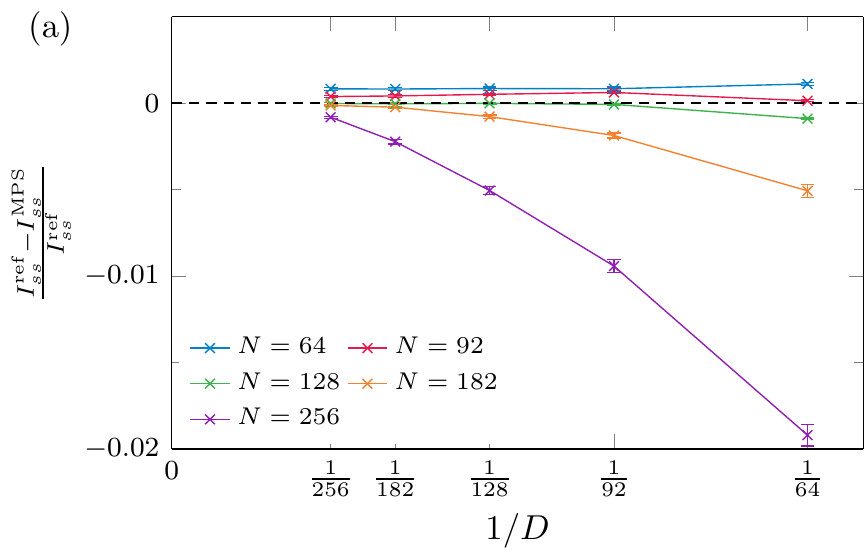}
\includegraphics[width=0.55\columnwidth]{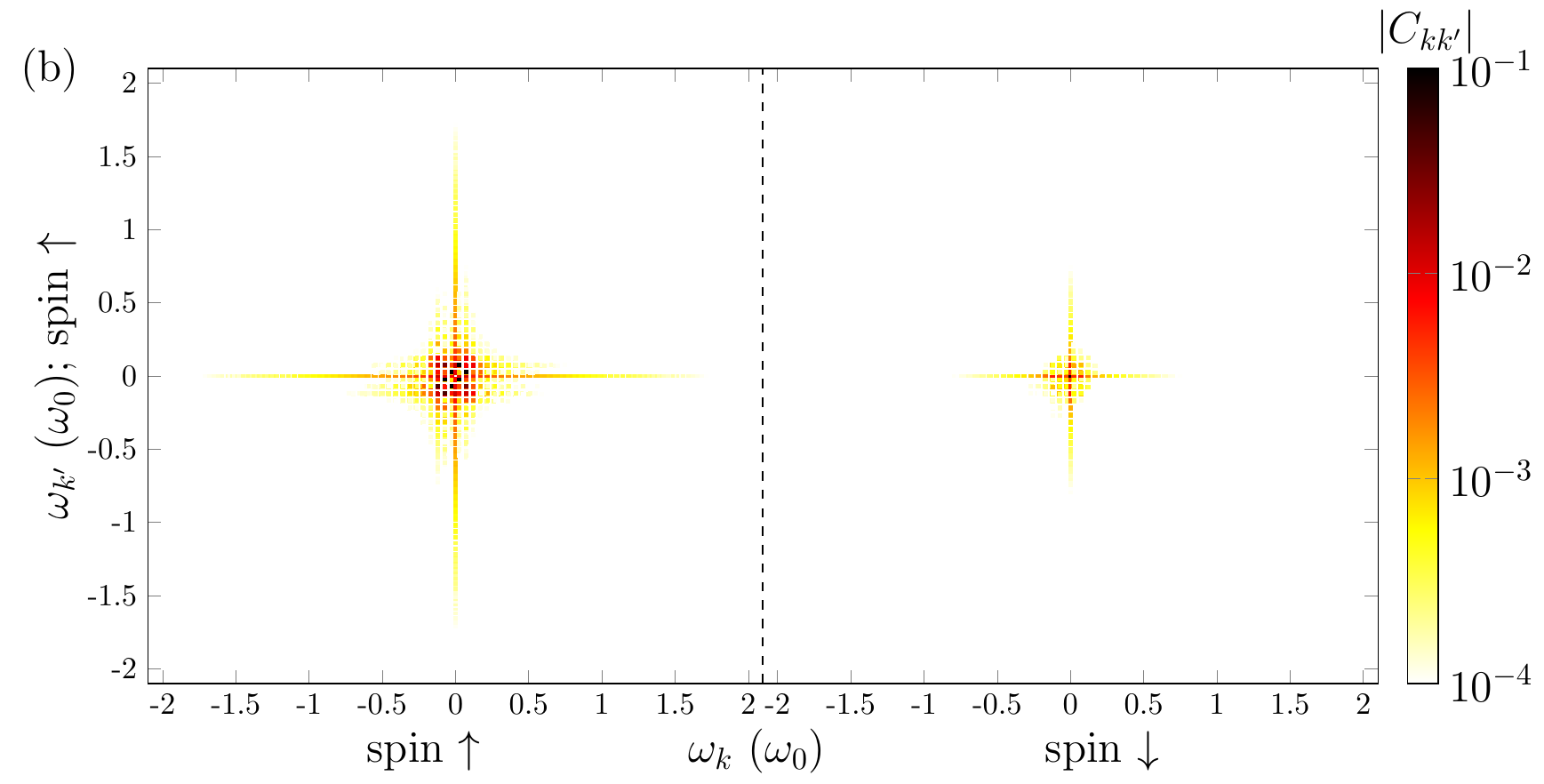}
\end{centering}
\vspace{-5pt}
\caption{{\bf Characterization of the many-body simulations.} The setup is the same as in Fig.~\ref{fig:anderson1} of the main text with $U=2$ and $\omega_S = -1$, which is the ``bifurcation'' point in Fig.~\ref{fig:anderson1}(d). 
(a) The relative difference of the steady-state current for a given point, $I^{\rm MPS}_{ss}$ at a particular $N$ and $D$, and a reference point, $I^{\rm ref}_{ss}$ at $N=128$ and $D=128$. For $N=128$ and smaller, the current is already converged to within $0.1$ \% for $D \gtrsim 92$. After this convergence, the smaller reservoir sizes have a gap between their currents and the reference current that represents the decrease in accuracy due to the finite reservoir size. For the larger reservoir sizes, the current estimate deviates even more, taking a larger $D$ to converge. This indicates that there is a trade-off between reservoir size and matrix product dimension when improving accuracy, as is the case for the non-interacting regime, see Fig.~\ref{fig:S1}. For a given error, it is best to choose a sufficient $N$ and increase $D$. As the desired error is decreases, though, a larger $N$ is necessary.
(b) Connected density-density correlation function $C_{k k'} = \langle n_k n_{k'} \rangle - \langle n_k \rangle \langle n_{k'} \rangle$ in the mixed basis at the latest time $t\omega_0 = N/2$ (here, $N=128$). Several features are clearly visible: 
(i) There are strong spin-up-spin-up correlations (the spin-down-spin-down correlations are identical) between the neighbouring modes with $\omega_k$ in the bias window, which is set at $\mu=0.2$  $\omega_0$  (left panel). In each spin channel the sites are ordered as in Eq.~\eqref{eq:MPOH} -- similarly as for $U=0$ discussed above.
(ii) At the same time, there are correlations between the spin channels (right panel), both at the impurity (placed at $\omega=0$) and the modes in/around the bias windows. This distribution of correlations corroborates the use of the $X$-lattice, Fig.~\ref{fig:anderson1}(a) of the main text, with the impurity put inside the bias windows and the spin-channels in proximity to each other in the tensor network structure.\label{fig:S4}}
\end{figure}

For the simulations in Fig.~\ref{fig:Inewmethod} of the main text, we set a threshold on the Schmidt values kept, $s_{min}$. The bond dimension is limited by a number of Schmidt values larger than 
$s_{min}$ and a maximal bond dimension $D$ -- whichever is smaller. We use $s_{min}=10^{-6}$ in most of the simulations, which we check is small enough not to influence the results. In the energy representation, the modes outside of the bias window ($-\mu/2$ to $\mu/2$) remain weakly entangled. For that reason, setting $s_{min}$ leads to a small bond dimension in that region and greatly speeds up the simulations, as indicated in the main text. Only the modes in the bias window are getting  entangled, and the precision of the simulation for longer times is ultimately controlled by $D$. 
We show typical behavior of the errors in Fig.~\ref{fig:S1} and provide some data on convergence of the conductance in Fig.~\ref{fig:anderson1}(d) of the main text in Fig.~\ref{fig:S4}.
We discuss further details of the simulations at the end of this Supplementary Material.

There are alternative setups/structures to handle the time dynamics in the energy basis. Multi-configuration time-dependent Hartree methods employ conditional states on the impurity~\cite{meyer_multi-configurational_1990,beck_multiconfiguration_2000}. This was recently employed with MPS for bosonic baths~\cite{kloss_multi-set_2018}. Thus, in addition to the physical structure of entanglement addressed here, there are also questions regarding the optimal implementation, which we will examine in a further contribution. We employ a standard MPS structure as we conjecture it will be the most scalable when going to larger-dimensional systems $\s$. 

{\em Effective entanglement entropy.} To facilitate the comparison between the computational cost of the spatial and mixed bases, we define an effective measure of entanglement of the lattice of a single channel, 
\begin{equation}
S_{\rm eff}= \ln \sqrt[\leftroot{-1}\uproot{2}\scriptstyle 3]{\frac{1}{L-1} \sum_n e^{3 S_n}},
\label{eq:Seff}
\end{equation}
where the sum is over all bipartite cuts (see Eq.~\eqref{eq:MPS} and comment below it for $S_n$), and $L-1$ is the total number of relevant cuts. The definition stems from the fact that maximal amount of describable entanglement scales logarithmically with the bond dimension $D_n$, and that the leading computational cost of simulation local step of the time evolution is proportional to $D_n^3$ (simplifying here that the neighboring bonds have the same $D_n$ and limiting ourselves to the 1D ordering). As such, equation \eqref{eq:Seff} incorporates that the required $D$ is inhomogeneous across the cuts and includes, heuristically, how the entanglement entropy contributes to computational cost.

{\it Comments on TDVP simulations.} The TDVP procedure, which we rely on to simulate the time evolution of MPS, is masterfully explained in Ref.~\onlinecite{haegeman_unifying_2016}. We briefly summarize it here in order to outline two variations which we employ in this article: combining 1-site and 2-site TDVP updates to, for efficiency, enlarge the bond dimension only when necessary and simulating the time evolution on the $X$-lattice in Fig.~\ref{fig:anderson1}(a) (or, more generally, on a tree).

Let's consider a quantum state represented as an MPS of length $L$,
\begin{equation}
\begin{aligned}
\includegraphics[width=0.75\columnwidth]{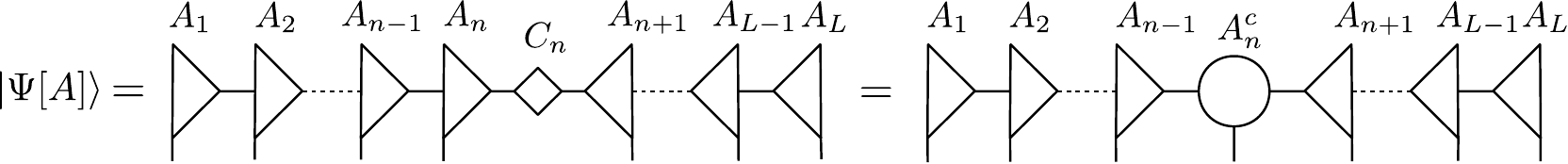}
\end{aligned},
\label{eq:MPS}
\end{equation}
with dangling legs corresponding to local physical degrees of freedom and connected lines corresponding to virtual degrees of freedom of the tensor network.
Above, we depict two representations of the same state in different mixed-canonical forms. All MPS tensors $A_i$ to the left (right) of the $n^{\mathrm{th}}$ bond are in left (right) canonical form~\cite{schollwock_density-matrix_2011} -- marked here using right-pointing (left-pointing) triangles. For instance, the entanglement of bipartite cuts appearing in Eq.~\eqref{eq:Seff} is fully encoded in the singular values of $C_n$ -- which we mark here as $\Lambda_{n,i}$ -- and reads $S_n = -2 \sum_{i=1}^{D_n} \Lambda_{n,i}^2\log_2 {\Lambda_{n,i}^2} $ (for a normalized state and bond dimension $D_n$ of the $n^{\mathrm{th}}$ cut). We also define $s_n = \min_{i} \Lambda_{n,i}$ as the smallest singular value for bond dimension $D_n$.

We can now consider the Hamiltonian $H$ and its expectation value in the state $|\Psi[A]\rangle$,
\begin{equation}
\begin{aligned}
\includegraphics[width=0.75\columnwidth]{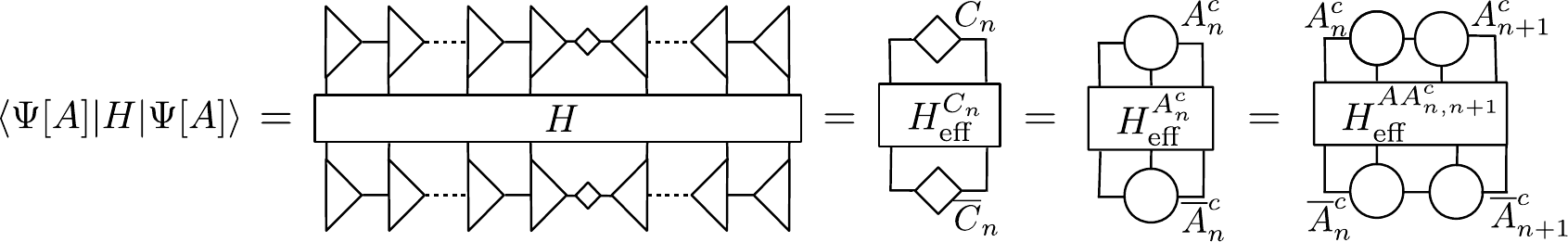}
\end{aligned}.
\end{equation}
The 0-site effective Hamiltonian $H_{\rm eff}^{C_n}$, related with the {central block} $C_n$, follows from the expectation value of $H$ in $|\Psi[A]\rangle$, which is calculated/contracted all the way except the contribution from $C_n$. All the MPS tensors which are contracted to form $H_{\rm eff}^{C_n}$ are in proper left and right canonical forms with respect to the position of the central block $C_n$. Similarly, one introduces the 1-site effective Hamiltonian $H_{\rm eff}^{A_n^c}$ related with the (central) MPS tensor $A_n^c$, and 2-site effective Hamiltonian $H_{\rm eff}^{AA^c_{n,n+1}}$ for two adjacent MPS tensors $A_n^c$ and $A_{n+1}^c$ blocked together.

In order to simulate the Schr\"odinger equation, TDVP projects the action of $H$ on $|\Psi[A]\rangle$ on the tangent space of $|\Psi[A]\rangle$~\cite{haegeman_time-dependent_2011}. To efficiently integrate it, the evolution operator is approximately decomposed into the set of local gates, which are used to update central blocks/sites~\cite{haegeman_unifying_2016}, 
\begin{eqnarray}
U^{[0]}_n(dt) C_n &=& \exp(- \imath dt H_{\rm eff}^{C_n}) C_n \rightarrow C_n  \\
U^{[1]}_n(dt) A^c_n &=& \exp(- \imath dt H_{\rm eff}^{A^c_n}) A^c_n \rightarrow A^c_n  \\
U^{[2]}_{n,n+1}(dt) A^c_n A^c_{n+1} &=& \exp(- \imath dt H_{\rm eff}^{AA^c_{n,n+1}}) A^c_n A^c_{n,n+1} \xrightarrow{\text{SVD}} A^c_n , A^c_{n+1} 
\end{eqnarray}
We note, again, that above the MPS is in a correct mixed canonical form with respect to the updated elements. In practice, one does not calculate the matrix representation of the effective Hamiltonian, but, for efficiency, employs a Krylov-based procedure (we use the method in Ref.~\cite{Niesen_Krylov}), which requires only the action of the effective Hamiltonian on a trial vector.  The latter can be efficiently calculated by combining smaller building blocks (environments) when $H$ is given as an MPO (or as a sum of local terms or MPOs).

Ref.~\onlinecite{haegeman_unifying_2016} discusses two main decompositions to simulate the time evolution,
\begin{equation}
    U(dt) \approx T^{[1]} (dt/2) T^{[1]*} (dt/2) \approx T^{[2]} (dt/2) T^{[2]*}(dt/2).
\end{equation}
For the 1-site TDVP scheme 
\begin{equation}
    T^{[1]}(y) = U^{[1]}_L(y) U^{[0]}_{L-1}(-y) U^{[1]}_{L-1}(y)  \ldots U^{[0]}_2(-y) U^{[1]}_{2}(y) U^{[0]}_{1}(-y) U^{[1]}_{1}(y),
\end{equation}
with central sites $A^c_n$ evolved forward in time, and central blocks $C_n$ evolved backward in time. $T^{[1]}$  constitute one sweep from left to right, where, before each local unitary update, MPS is put in a proper mixed canonical form. It is completed by its adjoint $T^{[1]*}$ with all the gates applied in the reverse order, i.e., a sweep from right to left, making it a $2^\mathrm{nd}$ order method in $dt$. It operates with fixed bond dimensions (at each cut) of the MPS.

Dynamical adjusting of the bond dimensions can be obtained by 2-site TDVP scheme
\begin{equation}
T^{[2]}(y) = U^{[2]}_{L-1,L}(y) U^{[1]}_{L-1}(-y) U^{[2]}_{L-2,L-1}(y)  \ldots U^{[1]}_3(-y) U^{[2]}_{2,3}(y) U^{[1]}_{2}(-y) U^{[2]}_{1,2}(y),
\end{equation}
(plus its adjoint $T^{[2]*}$ in the reverse order).  Now, the 2-site gate are evolved forward in time, and 1-site gate are evolved backward in time. In case of the 2-site update $U^{[2]}$, two adjacent MPS matrices are blocked together and subsequently split using a  singular value decomposition (SVD), truncating the virtual bond to given size/weights. It is, however, numerically significantly more costly both due to larger vectors appearing in the Krylov procedure and the additional SVD.

In this article, we employ a slight modification of the above procedures by combining the two schemes. The 2-site gates are employed only locally when both are necessary (all the Schmidt values of a given cut above a threshold $s_{min}$) and possible (bond dimension of a given cut is below the maximal $D$). Such an approach is consistent with the strongly inhomogeneous nature of the system we consider where entanglement is localized only in a part of the system, and the MPS bond dimensions between modes outside of the bias window can remain small.
To that end, it is sufficient to note how to transition between parts of $T^{[1]}(y)$ and $T^{[2]}(y)$ during a sweep to build $T^{[{\rm mixed}]}(y)$. If the $n$-th bond is enlarged and the next one is not, one gets $T^{[{\rm mixed}]}(y) = \ldots U^{[0]}_{n+1}(-y) U^{[2]}_{n,n+1}(y) \ldots$.
On the other hand, if $n^{\mathrm{th}}$ bond is not enlarged and the next one is, one has $T^{[{\rm mixed}]}(y) = \ldots U^{[2]}_{n+1,n+2}(y) U^{[0]}_{n}(-y) \ldots$. 
We perform the adjoint (sweep from right to left) using the exact reversal of the gates in $T^{[{\rm mixed}]}(y)$. We numerically observe, however, that finding new $T^{[{\rm mixed}]*}(y)$ based on the bond dimension/Schmidt weights does not reduce the order of the method in $dt$.
For clarity, below we collect the full procedure as a pseudocode \ref{alg:1}.

\begin{algorithm}
\begin{algorithmic}
\REQUIRE $|\Psi[A] \rangle$ in right canonical form; \\ minimal Schmidt values of all cuts $s_n$; \\ precomputed environments for calculation of $H_{\rm eff}$
\ENSURE $|\Psi[A]\rangle \leftarrow U(dt) |\Psi[A]\rangle$ in right canonical form; 
\STATE $C \leftarrow 1$
\STATE update\_two $\leftarrow$ {\bf false}
\FOR{$n=1$ {\bf to} $L$}
    \IF {{\bf not} update\_two}
        \STATE  $A^c \leftarrow C \cdot A_n$ 
        \IF {($s_n < s_{min}$) {\bf or} ($D_{n} \ge D_{max})$} \Comment{do not enlarge next bond dimension}
            \STATE $A^c \leftarrow U^{[1]}_n(dt/2) A^c$
            \STATE $A_n,\ C \leftarrow$ left-orthogonalize $A^c$
            \STATE update environments for calculation of $H_{\rm eff}$
            \STATE $C \leftarrow U^{[0]}_n(-dt/2) C$
        \ELSE
            \STATE update\_two $\leftarrow$ {\bf true}
        \ENDIF
    \ELSE
        \STATE $AA^c \leftarrow A^c \cdot A_n$
        \STATE $AA^c \leftarrow U^{[2]}_{n-1,n}(dt/2) AA^c$
        \STATE $A_{n-1},\ A^c \leftarrow $  left-orthogonalize and truncate $AA^c$ based on $D_{max}$ and $s_{min}$ (with some margin).
        \STATE update environments for calculation of  $H_{\rm eff}$
        \IF {($s_n < s_{min})$ {\bf or} ($D_n \ge D_{max}$) {\bf or} ($n = L$)} \Comment{do not enlarge next bond dimension}
            \STATE update\_two $\leftarrow$ {\bf false}
            \STATE $A_n,\ C \leftarrow$ left\_orthogonalize$(A^c)$
            \STATE  update environments for calculation of $H_{\rm eff}$
            \STATE $C \leftarrow U^{[0]}_{n}(-dt/2) C$
        \ELSE
            \STATE $A^c \leftarrow U^{[1]}_n(-dt/2) A^c$
        \ENDIF
    \ENDIF
\ENDFOR
\STATE sweep $n$ back from $L$ to $1$ additionally computing the minimal Schmidt values $s_n$.
\end{algorithmic}
\caption{Combining 1-site and 2-site TDVP sweep to locally enlarge the MPS bond dimension  based on a maximal bond dimension $D_{max}$ and Schmidt--value threshold $s_{min}$.
\label{alg:1}}
\end{algorithm}

Finally, the tree-tensor network ansatz corresponding to the X-shape lattice in Fig.~\ref{fig:anderson1}(a) of the main text is 
\begin{equation}
\begin{aligned}
\includegraphics[width=0.8\columnwidth]{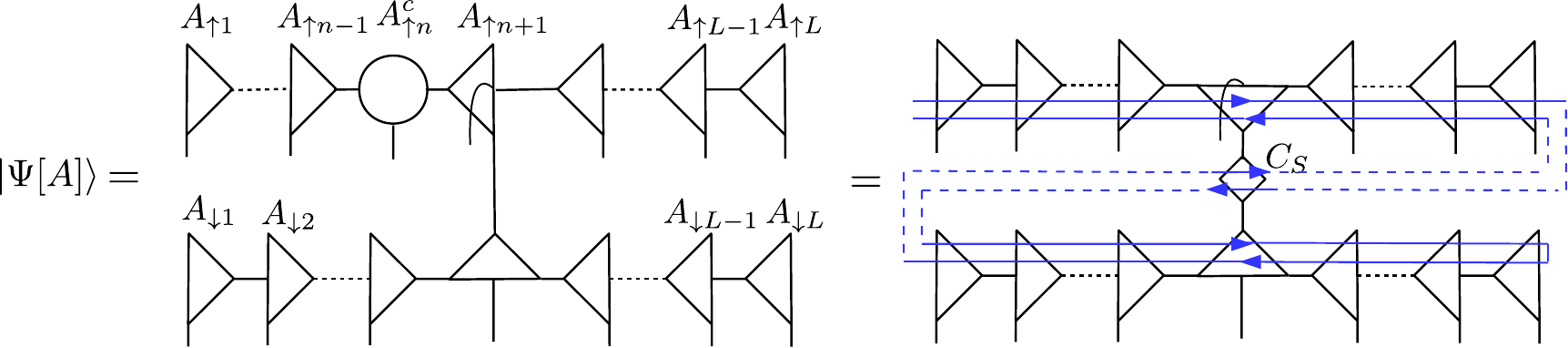}
\end{aligned}.
\label{eq:Xlattice}
\end{equation}
Again, we show two different mixed canonical representations of the same state with all the tensors to the left (right) of the central site [bond on the right-hand side] in the left (right) canonical form. On the right-hand side, we show the central block $C_S$ between the two spin channels, which in our case corresponds to the placement of the impurity $\mathcal{S}$ (hence the index).

The simulation of the time evolution is obtained using $U(dt) \approx T^{X}(dt/2) T^{X*}(dt/2)$. 
A one-way sweep is composed as $T^{X}(y) = T^{\uparrow [mixed]}(y) U^{[0]}_{\mathcal S} (-y) T^{\downarrow[mixed]}(y)$. 
Sweeps of the spin channels, $T^{\uparrow(\downarrow) [mixed]}(y)$, are done similarly as for the 1D chain above, and $U^{[0]}_{\mathcal S}$ describes the update of the central block $C_S$. If one wants to enlarge that bond, one can replace $U^{[0]}_{\mathcal S}(-y)$ with $U^{[1]}_{\downarrow \mathcal S}(-y) U^{[2]}_{\mathcal S}(y) U^{[1]}_{\uparrow \mathcal S}(-y)$, with $U^{[2]}_{\mathcal S}$ acting on two tensors connecting the spin channels, and $U^{[1]}_{\uparrow(\downarrow) \mathcal S}$ on each of them.
We depict the order of the full sweep -- combining to $U(dt)$ -- with blue arrows on the right-hand side of Eq.~\eqref{eq:Xlattice}. Its symmetric form ensures that it is second order in $dt$. The total Hamiltonian (which generates the $U$ gates) is treated as a sum of (the contributions coming from) MPOs for each channel and the interacting term, $U n_\downarrow n_\uparrow$, which is coupling the system modes placed next to each other in the $X$-lattice geometry.


\begin{thebibliography}{70}%
\makeatletter
\providecommand \@ifxundefined [1]{%
 \@ifx{#1\undefined}
}%
\providecommand \@ifnum [1]{%
 \ifnum #1\expandafter \@firstoftwo
 \else \expandafter \@secondoftwo
 \fi
}%
\providecommand \@ifx [1]{%
 \ifx #1\expandafter \@firstoftwo
 \else \expandafter \@secondoftwo
 \fi
}%
\providecommand \natexlab [1]{#1}%
\providecommand \enquote  [1]{``#1''}%
\providecommand \bibnamefont  [1]{#1}%
\providecommand \bibfnamefont [1]{#1}%
\providecommand \citenamefont [1]{#1}%
\providecommand \href@noop [0]{\@secondoftwo}%
\providecommand \href [0]{\begingroup \@sanitize@url \@href}%
\providecommand \@href[1]{\@@startlink{#1}\@@href}%
\providecommand \@@href[1]{\endgroup#1\@@endlink}%
\providecommand \@sanitize@url [0]{\catcode `\\12\catcode `\$12\catcode
  `\&12\catcode `\#12\catcode `\^12\catcode `\_12\catcode `\%12\relax}%
\providecommand \@@startlink[1]{}%
\providecommand \@@endlink[0]{}%
\providecommand \url  [0]{\begingroup\@sanitize@url \@url }%
\providecommand \@url [1]{\endgroup\@href {#1}{\urlprefix }}%
\providecommand \urlprefix  [0]{URL }%
\providecommand \Eprint [0]{\href }%
\providecommand \doibase [0]{http://dx.doi.org/}%
\providecommand \selectlanguage [0]{\@gobble}%
\providecommand \bibinfo  [0]{\@secondoftwo}%
\providecommand \bibfield  [0]{\@secondoftwo}%
\providecommand \translation [1]{[#1]}%
\providecommand \BibitemOpen [0]{}%
\providecommand \bibitemStop [0]{}%
\providecommand \bibitemNoStop [0]{.\EOS\space}%
\providecommand \EOS [0]{\spacefactor3000\relax}%
\providecommand \BibitemShut  [1]{\csname bibitem#1\endcsname}%
\let\auto@bib@innerbib\@empty
\bibitem [{\citenamefont {Orus}(2019)}]{orus_tensor_2018}%
  \BibitemOpen
  \bibfield  {author} {\bibinfo {author} {\bibfnamefont {R.}~\bibnamefont
  {Orus}},\ }\href {http://arxiv.org/abs/1812.04011} {\bibfield  {journal}
  {\bibinfo  {journal} {Nat. Rev. Phys.}\ }\textbf {\bibinfo {volume} {1}},\
  \bibinfo {pages} {538} (\bibinfo {year} {2019})}\BibitemShut {NoStop}%
\bibitem [{\citenamefont {Ran}\ \emph {et~al.}(2020)\citenamefont {Ran},
  \citenamefont {Tirrito}, \citenamefont {Peng}, \citenamefont {Chen},
  \citenamefont {Su},\ and\ \citenamefont {Lewenstein}}]{ran_review_2017}%
  \BibitemOpen
  \bibfield  {author} {\bibinfo {author} {\bibfnamefont {S.-J.}\ \bibnamefont
  {Ran}}, \bibinfo {author} {\bibfnamefont {E.}~\bibnamefont {Tirrito}},
  \bibinfo {author} {\bibfnamefont {C.}~\bibnamefont {Peng}}, \bibinfo {author}
  {\bibfnamefont {X.}~\bibnamefont {Chen}}, \bibinfo {author} {\bibfnamefont
  {G.}~\bibnamefont {Su}}, \ and\ \bibinfo {author} {\bibfnamefont
  {M.}~\bibnamefont {Lewenstein}},\ }\href
  {https://link.springer.com/book/10.1007%2F978-3-030-34489-4} {\emph {\bibinfo
  {title} {Tensor Network Contractions}}}\ (\bibinfo  {publisher} {Springer},\
  \bibinfo {year} {2020})\BibitemShut {NoStop}%
\bibitem [{\citenamefont {Orús}(2014)}]{orus_practical_2014}%
  \BibitemOpen
  \bibfield  {author} {\bibinfo {author} {\bibfnamefont {R.}~\bibnamefont
  {Orús}},\ }\href
  {http://www.sciencedirect.com/science/article/pii/S0003491614001596}
  {\bibfield  {journal} {\bibinfo  {journal} {Ann. Phys. (Amsterdam)}\ }\textbf
  {\bibinfo {volume} {349}},\ \bibinfo {pages} {117} (\bibinfo {year}
  {2014})}\BibitemShut {NoStop}%
\bibitem [{\citenamefont {Eisert}(2013)}]{eisert_entanglement_2013}%
  \BibitemOpen
  \bibfield  {author} {\bibinfo {author} {\bibfnamefont {J.}~\bibnamefont
  {Eisert}},\ }\href@noop {} {\bibfield  {journal} {\bibinfo  {journal} {Model.
  Simul.}\ }\textbf {\bibinfo {volume} {3}},\ \bibinfo {pages} {520} (\bibinfo
  {year} {2013})},\ \bibinfo {note}
  {\href{http://arxiv.org/abs/1308.3318}{arXiv:1308.3318}}\BibitemShut
  {NoStop}%
\bibitem [{\citenamefont
  {Schollw\"ock}(2011)}]{schollwock_density-matrix_2011}%
  \BibitemOpen
  \bibfield  {author} {\bibinfo {author} {\bibfnamefont {U.}~\bibnamefont
  {Schollw\"ock}},\ }\href
  {https://www.sciencedirect.com/science/article/pii/S0003491610001752}
  {\bibfield  {journal} {\bibinfo  {journal} {Ann. Phys. (Amsterdam)}\ }\textbf
  {\bibinfo {volume} {326}},\ \bibinfo {pages} {96} (\bibinfo {year}
  {2011})}\BibitemShut {NoStop}%
\bibitem [{\citenamefont {Verstraete}\ \emph {et~al.}(2008)\citenamefont
  {Verstraete}, \citenamefont {Murg},\ and\ \citenamefont
  {Cirac}}]{verstraete_review_2008}%
  \BibitemOpen
  \bibfield  {author} {\bibinfo {author} {\bibfnamefont {F.}~\bibnamefont
  {Verstraete}}, \bibinfo {author} {\bibfnamefont {V.}~\bibnamefont {Murg}}, \
  and\ \bibinfo {author} {\bibfnamefont {J.}~\bibnamefont {Cirac}},\ }\href
  {\doibase 10.1080/14789940801912366} {\bibfield  {journal} {\bibinfo
  {journal} {Adv. Phys.}\ }\textbf {\bibinfo {volume} {57}},\ \bibinfo {pages}
  {143} (\bibinfo {year} {2008})}\BibitemShut {NoStop}%
\bibitem [{\citenamefont {Kaufman}\ \emph {et~al.}(2016)\citenamefont
  {Kaufman}, \citenamefont {Tai}, \citenamefont {Lukin}, \citenamefont
  {Rispoli}, \citenamefont {Schittko}, \citenamefont {Preiss},\ and\
  \citenamefont {Greiner}}]{kaufman_quantum_2016}%
  \BibitemOpen
  \bibfield  {author} {\bibinfo {author} {\bibfnamefont {A.~M.}\ \bibnamefont
  {Kaufman}}, \bibinfo {author} {\bibfnamefont {M.~E.}\ \bibnamefont {Tai}},
  \bibinfo {author} {\bibfnamefont {A.}~\bibnamefont {Lukin}}, \bibinfo
  {author} {\bibfnamefont {M.}~\bibnamefont {Rispoli}}, \bibinfo {author}
  {\bibfnamefont {R.}~\bibnamefont {Schittko}}, \bibinfo {author}
  {\bibfnamefont {P.~M.}\ \bibnamefont {Preiss}}, \ and\ \bibinfo {author}
  {\bibfnamefont {M.}~\bibnamefont {Greiner}},\ }\href
  {http://science.sciencemag.org/content/353/6301/794} {\bibfield  {journal}
  {\bibinfo  {journal} {Science}\ }\textbf {\bibinfo {volume} {353}},\ \bibinfo
  {pages} {794} (\bibinfo {year} {2016})}\BibitemShut {NoStop}%
\bibitem [{\citenamefont {Alba}\ and\ \citenamefont
  {Calabrese}(2017)}]{alba_entanglement_2017}%
  \BibitemOpen
  \bibfield  {author} {\bibinfo {author} {\bibfnamefont {V.}~\bibnamefont
  {Alba}}\ and\ \bibinfo {author} {\bibfnamefont {P.}~\bibnamefont
  {Calabrese}},\ }\href {https://www.pnas.org/content/114/30/7947} {\bibfield
  {journal} {\bibinfo  {journal} {Proc. Natl. Acad. Sci. U.S.A.}\ }\textbf
  {\bibinfo {volume} {114}},\ \bibinfo {pages} {7947} (\bibinfo {year}
  {2017})}\BibitemShut {NoStop}%
\bibitem [{\citenamefont {Liu}\ and\ \citenamefont
  {Suh}(2014)}]{liu_entanglement_2014}%
  \BibitemOpen
  \bibfield  {author} {\bibinfo {author} {\bibfnamefont {H.}~\bibnamefont
  {Liu}}\ and\ \bibinfo {author} {\bibfnamefont {S.~J.}\ \bibnamefont {Suh}},\
  }\href {https://link.aps.org/doi/10.1103/PhysRevLett.112.011601} {\bibfield
  {journal} {\bibinfo  {journal} {Phys. Rev. Lett.}\ }\textbf {\bibinfo
  {volume} {112}},\ \bibinfo {pages} {011601} (\bibinfo {year}
  {2014})}\BibitemShut {NoStop}%
\bibitem [{\citenamefont {Kim}\ and\ \citenamefont
  {Huse}(2013)}]{kim_ballistic_2013}%
  \BibitemOpen
  \bibfield  {author} {\bibinfo {author} {\bibfnamefont {H.}~\bibnamefont
  {Kim}}\ and\ \bibinfo {author} {\bibfnamefont {D.~A.}\ \bibnamefont {Huse}},\
  }\href {https://link.aps.org/doi/10.1103/PhysRevLett.111.127205} {\bibfield
  {journal} {\bibinfo  {journal} {Phys. Rev. Lett.}\ }\textbf {\bibinfo
  {volume} {111}},\ \bibinfo {pages} {127205} (\bibinfo {year}
  {2013})}\BibitemShut {NoStop}%
\bibitem [{\citenamefont {Schachenmayer}\ \emph {et~al.}(2013)\citenamefont
  {Schachenmayer}, \citenamefont {Lanyon}, \citenamefont {Roos},\ and\
  \citenamefont {Daley}}]{schachenmayer_entanglement_2013}%
  \BibitemOpen
  \bibfield  {author} {\bibinfo {author} {\bibfnamefont {J.}~\bibnamefont
  {Schachenmayer}}, \bibinfo {author} {\bibfnamefont {B.~P.}\ \bibnamefont
  {Lanyon}}, \bibinfo {author} {\bibfnamefont {C.~F.}\ \bibnamefont {Roos}}, \
  and\ \bibinfo {author} {\bibfnamefont {A.~J.}\ \bibnamefont {Daley}},\ }\href
  {https://link.aps.org/doi/10.1103/PhysRevX.3.031015} {\bibfield  {journal}
  {\bibinfo  {journal} {Phys. Rev. X}\ }\textbf {\bibinfo {volume} {3}},\
  \bibinfo {pages} {031015} (\bibinfo {year} {2013})}\BibitemShut {NoStop}%
\bibitem [{\citenamefont {Schuch}\ \emph {et~al.}(2008)\citenamefont {Schuch},
  \citenamefont {Wolf}, \citenamefont {Vollbrecht},\ and\ \citenamefont
  {Cirac}}]{schuch_entropy_2008}%
  \BibitemOpen
  \bibfield  {author} {\bibinfo {author} {\bibfnamefont {N.}~\bibnamefont
  {Schuch}}, \bibinfo {author} {\bibfnamefont {M.~M.}\ \bibnamefont {Wolf}},
  \bibinfo {author} {\bibfnamefont {K.~G.~H.}\ \bibnamefont {Vollbrecht}}, \
  and\ \bibinfo {author} {\bibfnamefont {J.~I.}\ \bibnamefont {Cirac}},\ }\href
  {https://doi.org/10.1088%2F1367-2630%2F10%2F3%2F033032} {\bibfield  {journal}
  {\bibinfo  {journal} {New J. Phys.}\ }\textbf {\bibinfo {volume} {10}},\
  \bibinfo {pages} {033032} (\bibinfo {year} {2008})}\BibitemShut {NoStop}%
\bibitem [{\citenamefont {Calabrese}\ and\ \citenamefont
  {Cardy}(2005)}]{calabrese_evolution_2005}%
  \BibitemOpen
  \bibfield  {author} {\bibinfo {author} {\bibfnamefont {P.}~\bibnamefont
  {Calabrese}}\ and\ \bibinfo {author} {\bibfnamefont {J.}~\bibnamefont
  {Cardy}},\ }\href {https://doi.org/10.1088%2F1742-5468%2F2005%2F04%2Fp04010}
  {\bibfield  {journal} {\bibinfo  {journal} {J. Stat. Mech.: Theory Exp.}\ ,\
  \bibinfo {pages} {P04010}} (\bibinfo {year} {2005})}\BibitemShut {NoStop}%
\bibitem [{\citenamefont {White}\ \emph {et~al.}(2018)\citenamefont {White},
  \citenamefont {Zaletel}, \citenamefont {Mong},\ and\ \citenamefont
  {Refael}}]{Refael_2018}%
  \BibitemOpen
  \bibfield  {author} {\bibinfo {author} {\bibfnamefont {C.~D.}\ \bibnamefont
  {White}}, \bibinfo {author} {\bibfnamefont {M.}~\bibnamefont {Zaletel}},
  \bibinfo {author} {\bibfnamefont {R.~S.~K.}\ \bibnamefont {Mong}}, \ and\
  \bibinfo {author} {\bibfnamefont {G.}~\bibnamefont {Refael}},\ }\href
  {\doibase 10.1103/PhysRevB.97.035127} {\bibfield  {journal} {\bibinfo
  {journal} {Phys. Rev. B}\ }\textbf {\bibinfo {volume} {97}},\ \bibinfo
  {pages} {035127} (\bibinfo {year} {2018})}\BibitemShut {NoStop}%
\bibitem [{\citenamefont {{Leviatan}}\ \emph {et~al.}(2017)\citenamefont
  {{Leviatan}}, \citenamefont {{Pollmann}}, \citenamefont {{Bardarson}},
  \citenamefont {{Huse}},\ and\ \citenamefont {{Altman}}}]{Altman2017}%
  \BibitemOpen
  \bibfield  {author} {\bibinfo {author} {\bibfnamefont {E.}~\bibnamefont
  {{Leviatan}}}, \bibinfo {author} {\bibfnamefont {F.}~\bibnamefont
  {{Pollmann}}}, \bibinfo {author} {\bibfnamefont {J.~H.}\ \bibnamefont
  {{Bardarson}}}, \bibinfo {author} {\bibfnamefont {D.~A.}\ \bibnamefont
  {{Huse}}}, \ and\ \bibinfo {author} {\bibfnamefont {E.}~\bibnamefont
  {{Altman}}},\ }\href {http://arxiv.org/abs/1702.08894} {\bibfield  {journal}
  {\bibinfo  {journal} {arXiv:1702.08894}\ } (\bibinfo {year}
  {2017})}\BibitemShut {NoStop}%
\bibitem [{\citenamefont {Surace}\ \emph {et~al.}(2019)\citenamefont {Surace},
  \citenamefont {Piani},\ and\ \citenamefont {Tagliacozzo}}]{Luca_2018}%
  \BibitemOpen
  \bibfield  {author} {\bibinfo {author} {\bibfnamefont {J.}~\bibnamefont
  {Surace}}, \bibinfo {author} {\bibfnamefont {M.}~\bibnamefont {Piani}}, \
  and\ \bibinfo {author} {\bibfnamefont {L.}~\bibnamefont {Tagliacozzo}},\
  }\href {\doibase 10.1103/PhysRevB.99.235115} {\bibfield  {journal} {\bibinfo
  {journal} {Phys. Rev. B}\ }\textbf {\bibinfo {volume} {99}},\ \bibinfo
  {pages} {235115} (\bibinfo {year} {2019})}\BibitemShut {NoStop}%
\bibitem [{\citenamefont {Beenakker}(2006)}]{beenakker_electron-hole_2006}%
  \BibitemOpen
  \bibfield  {author} {\bibinfo {author} {\bibfnamefont {C.~W.~J.}\
  \bibnamefont {Beenakker}},\ }in\ \href
  {https://doi.org/10.3254/978-1-61499-018-5-307} {\emph {\bibinfo {booktitle}
  {Proc. {Int}. {School} {Phys}. {E}. {Fermi}}}},\ Vol.\ \bibinfo {volume}
  {162}\ (\bibinfo  {publisher} {IOS Press, Amsterdam},\ \bibinfo {year}
  {2006})\ pp.\ \bibinfo {pages} {307--347}\BibitemShut {NoStop}%
\bibitem [{\citenamefont {Klich}\ and\ \citenamefont
  {Levitov}(2009)}]{klich_quantum_2009}%
  \BibitemOpen
  \bibfield  {author} {\bibinfo {author} {\bibfnamefont {I.}~\bibnamefont
  {Klich}}\ and\ \bibinfo {author} {\bibfnamefont {L.}~\bibnamefont
  {Levitov}},\ }\href {https://link.aps.org/doi/10.1103/PhysRevLett.102.100502}
  {\bibfield  {journal} {\bibinfo  {journal} {Phys. Rev. Lett.}\ }\textbf
  {\bibinfo {volume} {102}},\ \bibinfo {pages} {100502} (\bibinfo {year}
  {2009})}\BibitemShut {NoStop}%
\bibitem [{\citenamefont {Levitov}\ and\ \citenamefont
  {Lesovik}(1993)}]{levitov_charge_1993}%
  \BibitemOpen
  \bibfield  {author} {\bibinfo {author} {\bibfnamefont {L.~S.}\ \bibnamefont
  {Levitov}}\ and\ \bibinfo {author} {\bibfnamefont {G.~B.}\ \bibnamefont
  {Lesovik}},\ }\href
  {http://www.jetpletters.ac.ru/ps/1186/article_17907.shtml} {\bibfield
  {journal} {\bibinfo  {journal} {JETP Letters}\ }\textbf {\bibinfo {volume}
  {58}},\ \bibinfo {pages} {230} (\bibinfo {year} {1993})}\BibitemShut
  {NoStop}%
\bibitem [{\citenamefont {Chien}\ \emph {et~al.}(2014)\citenamefont {Chien},
  \citenamefont {Di~Ventra},\ and\ \citenamefont
  {Zwolak}}]{chien_landauer_2014}%
  \BibitemOpen
  \bibfield  {author} {\bibinfo {author} {\bibfnamefont {C.-C.}\ \bibnamefont
  {Chien}}, \bibinfo {author} {\bibfnamefont {M.}~\bibnamefont {Di~Ventra}}, \
  and\ \bibinfo {author} {\bibfnamefont {M.}~\bibnamefont {Zwolak}},\ }\href
  {\doibase 10.1103/PhysRevA.90.023624} {\bibfield  {journal} {\bibinfo
  {journal} {Phys. Rev. A}\ }\textbf {\bibinfo {volume} {90}},\ \bibinfo
  {pages} {023624} (\bibinfo {year} {2014})}\BibitemShut {NoStop}%
\bibitem [{\citenamefont {Cazalilla}\ and\ \citenamefont
  {Marston}(2002)}]{cazalilla_time-dependent_2002}%
  \BibitemOpen
  \bibfield  {author} {\bibinfo {author} {\bibfnamefont {M.~A.}\ \bibnamefont
  {Cazalilla}}\ and\ \bibinfo {author} {\bibfnamefont {J.~B.}\ \bibnamefont
  {Marston}},\ }\href {https://link.aps.org/doi/10.1103/PhysRevLett.88.256403}
  {\bibfield  {journal} {\bibinfo  {journal} {Phys. Rev. Lett.}\ }\textbf
  {\bibinfo {volume} {88}},\ \bibinfo {pages} {256403} (\bibinfo {year}
  {2002})}\BibitemShut {NoStop}%
\bibitem [{\citenamefont {Zwolak}\ and\ \citenamefont
  {Vidal}(2004)}]{zwolak_mixed-state_2004}%
  \BibitemOpen
  \bibfield  {author} {\bibinfo {author} {\bibfnamefont {M.}~\bibnamefont
  {Zwolak}}\ and\ \bibinfo {author} {\bibfnamefont {G.}~\bibnamefont {Vidal}},\
  }\href {\doibase 10.1103/PhysRevLett.93.207205} {\bibfield  {journal}
  {\bibinfo  {journal} {Phys. Rev. Lett.}\ }\textbf {\bibinfo {volume} {93}},\
  \bibinfo {pages} {207205} (\bibinfo {year} {2004})}\BibitemShut {NoStop}%
\bibitem [{\citenamefont {Gobert}\ \emph {et~al.}(2005)\citenamefont {Gobert},
  \citenamefont {Kollath}, \citenamefont {Schollw\"ock},\ and\ \citenamefont
  {Schütz}}]{gobert_real-time_2005}%
  \BibitemOpen
  \bibfield  {author} {\bibinfo {author} {\bibfnamefont {D.}~\bibnamefont
  {Gobert}}, \bibinfo {author} {\bibfnamefont {C.}~\bibnamefont {Kollath}},
  \bibinfo {author} {\bibfnamefont {U.}~\bibnamefont {Schollw\"ock}}, \ and\
  \bibinfo {author} {\bibfnamefont {G.}~\bibnamefont {Schütz}},\ }\href
  {https://link.aps.org/doi/10.1103/PhysRevE.71.036102} {\bibfield  {journal}
  {\bibinfo  {journal} {Phys. Rev. E}\ }\textbf {\bibinfo {volume} {71}},\
  \bibinfo {pages} {036102} (\bibinfo {year} {2005})}\BibitemShut {NoStop}%
\bibitem [{\citenamefont {Schneider}\ and\ \citenamefont
  {Schmitteckert}(2006)}]{schneider_conductance_2006}%
  \BibitemOpen
  \bibfield  {author} {\bibinfo {author} {\bibfnamefont {G.}~\bibnamefont
  {Schneider}}\ and\ \bibinfo {author} {\bibfnamefont {P.}~\bibnamefont
  {Schmitteckert}},\ }\href {http://arxiv.org/abs/cond-mat/0601389} {\bibfield
  {journal} {\bibinfo  {journal} {arXiv:cond-mat/0601389}\ } (\bibinfo {year}
  {2006})}\BibitemShut {NoStop}%
\bibitem [{\citenamefont {Schmitteckert}\ and\ \citenamefont
  {Schneider}(2006)}]{Schmitteckert06-1}%
  \BibitemOpen
  \bibfield  {author} {\bibinfo {author} {\bibfnamefont {P.}~\bibnamefont
  {Schmitteckert}}\ and\ \bibinfo {author} {\bibfnamefont {G.}~\bibnamefont
  {Schneider}},\ }in\ \href@noop {} {\emph {\bibinfo {booktitle} {High
  Performance Computing in Science and Engineering}}},\ \bibinfo {editor}
  {edited by\ \bibinfo {editor} {\bibfnamefont {W.~E.}\ \bibnamefont {Nagel}},
  \bibinfo {editor} {\bibfnamefont {W.}~\bibnamefont {J\"ager}}, \ and\
  \bibinfo {editor} {\bibfnamefont {M.}~\bibnamefont {Resch}}}\ (\bibinfo
  {publisher} {Springer},\ \bibinfo {address} {Berlin},\ \bibinfo {year}
  {2006})\ pp.\ \bibinfo {pages} {113--126}\BibitemShut {NoStop}%
\bibitem [{\citenamefont {Al-Hassanieh}\ \emph {et~al.}(2006)\citenamefont
  {Al-Hassanieh}, \citenamefont {Feiguin}, \citenamefont {Riera}, \citenamefont
  {Büsser},\ and\ \citenamefont {Dagotto}}]{al-hassanieh_adaptive_2006}%
  \BibitemOpen
  \bibfield  {author} {\bibinfo {author} {\bibfnamefont {K.~A.}\ \bibnamefont
  {Al-Hassanieh}}, \bibinfo {author} {\bibfnamefont {A.~E.}\ \bibnamefont
  {Feiguin}}, \bibinfo {author} {\bibfnamefont {J.~A.}\ \bibnamefont {Riera}},
  \bibinfo {author} {\bibfnamefont {C.~A.}\ \bibnamefont {Büsser}}, \ and\
  \bibinfo {author} {\bibfnamefont {E.}~\bibnamefont {Dagotto}},\ }\href
  {https://link.aps.org/doi/10.1103/PhysRevB.73.195304} {\bibfield  {journal}
  {\bibinfo  {journal} {Phys. Rev. B}\ }\textbf {\bibinfo {volume} {73}},\
  \bibinfo {pages} {195304} (\bibinfo {year} {2006})}\BibitemShut {NoStop}%
\bibitem [{\citenamefont {Dias~da Silva}\ \emph {et~al.}(2008)\citenamefont
  {Dias~da Silva}, \citenamefont {Heidrich-Meisner}, \citenamefont {Feiguin},
  \citenamefont {Büsser}, \citenamefont {Martins}, \citenamefont {Anda},\ and\
  \citenamefont {Dagotto}}]{dias_da_silva_transport_2008}%
  \BibitemOpen
  \bibfield  {author} {\bibinfo {author} {\bibfnamefont {L.~G. G.~V.}\
  \bibnamefont {Dias~da Silva}}, \bibinfo {author} {\bibfnamefont
  {F.}~\bibnamefont {Heidrich-Meisner}}, \bibinfo {author} {\bibfnamefont
  {A.~E.}\ \bibnamefont {Feiguin}}, \bibinfo {author} {\bibfnamefont {C.~A.}\
  \bibnamefont {Büsser}}, \bibinfo {author} {\bibfnamefont {G.~B.}\
  \bibnamefont {Martins}}, \bibinfo {author} {\bibfnamefont {E.~V.}\
  \bibnamefont {Anda}}, \ and\ \bibinfo {author} {\bibfnamefont
  {E.}~\bibnamefont {Dagotto}},\ }\href
  {https://link.aps.org/doi/10.1103/PhysRevB.78.195317} {\bibfield  {journal}
  {\bibinfo  {journal} {Phys. Rev. B}\ }\textbf {\bibinfo {volume} {78}},\
  \bibinfo {pages} {195317} (\bibinfo {year} {2008})}\BibitemShut {NoStop}%
\bibitem [{\citenamefont {Heidrich-Meisner}\ \emph {et~al.}(2009)\citenamefont
  {Heidrich-Meisner}, \citenamefont {Feiguin},\ and\ \citenamefont
  {Dagotto}}]{heidrich-meisner_real-time_2009}%
  \BibitemOpen
  \bibfield  {author} {\bibinfo {author} {\bibfnamefont {F.}~\bibnamefont
  {Heidrich-Meisner}}, \bibinfo {author} {\bibfnamefont {A.~E.}\ \bibnamefont
  {Feiguin}}, \ and\ \bibinfo {author} {\bibfnamefont {E.}~\bibnamefont
  {Dagotto}},\ }\href {https://link.aps.org/doi/10.1103/PhysRevB.79.235336}
  {\bibfield  {journal} {\bibinfo  {journal} {Phys. Rev. B}\ }\textbf {\bibinfo
  {volume} {79}},\ \bibinfo {pages} {235336} (\bibinfo {year}
  {2009})}\BibitemShut {NoStop}%
\bibitem [{\citenamefont {Branschädel}\ \emph {et~al.}(2010)\citenamefont
  {Branschädel}, \citenamefont {Schneider},\ and\ \citenamefont
  {Schmitteckert}}]{branschadel_conductance_2010}%
  \BibitemOpen
  \bibfield  {author} {\bibinfo {author} {\bibfnamefont {A.}~\bibnamefont
  {Branschädel}}, \bibinfo {author} {\bibfnamefont {G.}~\bibnamefont
  {Schneider}}, \ and\ \bibinfo {author} {\bibfnamefont {P.}~\bibnamefont
  {Schmitteckert}},\ }\href
  {https://onlinelibrary.wiley.com/doi/abs/10.1002/andp.201000017} {\bibfield
  {journal} {\bibinfo  {journal} {Ann. Phys. (Berlin)}\ }\textbf {\bibinfo
  {volume} {522}},\ \bibinfo {pages} {657} (\bibinfo {year}
  {2010})}\BibitemShut {NoStop}%
\bibitem [{\citenamefont {Chien}\ \emph {et~al.}(2013)\citenamefont {Chien},
  \citenamefont {Gruss}, \citenamefont {Di~Ventra},\ and\ \citenamefont
  {Zwolak}}]{chien_interaction-induced_2013}%
  \BibitemOpen
  \bibfield  {author} {\bibinfo {author} {\bibfnamefont {C.-C.}\ \bibnamefont
  {Chien}}, \bibinfo {author} {\bibfnamefont {D.}~\bibnamefont {Gruss}},
  \bibinfo {author} {\bibfnamefont {M.}~\bibnamefont {Di~Ventra}}, \ and\
  \bibinfo {author} {\bibfnamefont {M.}~\bibnamefont {Zwolak}},\ }\href
  {http://arxiv.org/abs/1203.5094} {\bibfield  {journal} {\bibinfo  {journal}
  {New J. Phys.}\ }\textbf {\bibinfo {volume} {15}},\ \bibinfo {pages} {063026}
  (\bibinfo {year} {2013})}\BibitemShut {NoStop}%
\bibitem [{\citenamefont {Gruss}\ \emph {et~al.}(2018)\citenamefont {Gruss},
  \citenamefont {Chien}, \citenamefont {Barreiro}, \citenamefont {Ventra},\
  and\ \citenamefont {Zwolak}}]{gruss_energy-resolved_2018}%
  \BibitemOpen
  \bibfield  {author} {\bibinfo {author} {\bibfnamefont {D.}~\bibnamefont
  {Gruss}}, \bibinfo {author} {\bibfnamefont {C.-C.}\ \bibnamefont {Chien}},
  \bibinfo {author} {\bibfnamefont {J.~T.}\ \bibnamefont {Barreiro}}, \bibinfo
  {author} {\bibfnamefont {M.~D.}\ \bibnamefont {Ventra}}, \ and\ \bibinfo
  {author} {\bibfnamefont {M.}~\bibnamefont {Zwolak}},\ }\href
  {http://stacks.iop.org/1367-2630/20/i=11/a=115005} {\bibfield  {journal}
  {\bibinfo  {journal} {New J. Phys.}\ }\textbf {\bibinfo {volume} {20}},\
  \bibinfo {pages} {115005} (\bibinfo {year} {2018})}\BibitemShut {NoStop}%
\bibitem [{\citenamefont {Bohr}\ \emph {et~al.}(2006)\citenamefont {Bohr},
  \citenamefont {Schmitteckert},\ and\ \citenamefont
  {W\"olfle}}]{bohr_dmrg_2006}%
  \BibitemOpen
  \bibfield  {author} {\bibinfo {author} {\bibfnamefont {D.}~\bibnamefont
  {Bohr}}, \bibinfo {author} {\bibfnamefont {P.}~\bibnamefont {Schmitteckert}},
  \ and\ \bibinfo {author} {\bibfnamefont {P.}~\bibnamefont {W\"olfle}},\
  }\href
  {https://epljournal.edpsciences.org/articles/epl/abs/2006/02/epl9125/epl9125.html}
  {\bibfield  {journal} {\bibinfo  {journal} {Europhys. Lett.}\ }\textbf
  {\bibinfo {volume} {73}},\ \bibinfo {pages} {246} (\bibinfo {year}
  {2006})}\BibitemShut {NoStop}%
\bibitem [{\citenamefont {Bohr}\ and\ \citenamefont
  {Schmitteckert}(2007)}]{bohr_strong_2007}%
  \BibitemOpen
  \bibfield  {author} {\bibinfo {author} {\bibfnamefont {D.}~\bibnamefont
  {Bohr}}\ and\ \bibinfo {author} {\bibfnamefont {P.}~\bibnamefont
  {Schmitteckert}},\ }\href
  {https://link.aps.org/doi/10.1103/PhysRevB.75.241103} {\bibfield  {journal}
  {\bibinfo  {journal} {Phys. Rev. B}\ }\textbf {\bibinfo {volume} {75}},\
  \bibinfo {pages} {241103} (\bibinfo {year} {2007})}\BibitemShut {NoStop}%
\bibitem [{\citenamefont {Wolf}\ \emph {et~al.}(2014)\citenamefont {Wolf},
  \citenamefont {McCulloch},\ and\ \citenamefont
  {Schollw\"ock}}]{wolf_solving_2014}%
  \BibitemOpen
  \bibfield  {author} {\bibinfo {author} {\bibfnamefont {F.~A.}\ \bibnamefont
  {Wolf}}, \bibinfo {author} {\bibfnamefont {I.~P.}\ \bibnamefont {McCulloch}},
  \ and\ \bibinfo {author} {\bibfnamefont {U.}~\bibnamefont {Schollw\"ock}},\
  }\href {https://link.aps.org/doi/10.1103/PhysRevB.90.235131} {\bibfield
  {journal} {\bibinfo  {journal} {Phys. Rev. B}\ }\textbf {\bibinfo {volume}
  {90}},\ \bibinfo {pages} {235131} (\bibinfo {year} {2014})}\BibitemShut
  {NoStop}%
\bibitem [{\citenamefont {He}\ and\ \citenamefont
  {Millis}(2017)}]{he_entanglement_2017}%
  \BibitemOpen
  \bibfield  {author} {\bibinfo {author} {\bibfnamefont {Z.}~\bibnamefont
  {He}}\ and\ \bibinfo {author} {\bibfnamefont {A.~J.}\ \bibnamefont
  {Millis}},\ }\href {https://link.aps.org/doi/10.1103/PhysRevB.96.085107}
  {\bibfield  {journal} {\bibinfo  {journal} {Phys. Rev. B}\ }\textbf {\bibinfo
  {volume} {96}},\ \bibinfo {pages} {085107} (\bibinfo {year}
  {2017})}\BibitemShut {NoStop}%
\bibitem [{Sup()}]{Supplementary}%
  \BibitemOpen
  \href@noop {} {}\bibinfo {note} {{See the Supplementary Material, which
  includes
  Refs.~\cite{jauho_time-dependent_1994,caroli_direct_1971,Wilson75-1,bulla_numerical_2008,vidal_efficient_2003,*vidal_efficient_2004,schmitteckert_nonequilibrium_2004,haegeman_time-dependent_2011,Niesen_Krylov,zwolak_communication:_2018,meyer_multi-configurational_1990,beck_multiconfiguration_2000,kloss_multi-set_2018},
  definitions of all operators, further evidence pertaining to the entanglement
  structure, scaling of errors in mixed-basis simulations, and additional
  details of the numerics.}}\BibitemShut {Stop}%
\bibitem [{\citenamefont {Verstraete}\ and\ \citenamefont
  {Cirac}(2006)}]{verstraete_faithfully_2006}%
  \BibitemOpen
  \bibfield  {author} {\bibinfo {author} {\bibfnamefont {F.}~\bibnamefont
  {Verstraete}}\ and\ \bibinfo {author} {\bibfnamefont {J.~I.}\ \bibnamefont
  {Cirac}},\ }\href {\doibase 10.1103/PhysRevB.73.094423} {\bibfield  {journal}
  {\bibinfo  {journal} {Phys. Rev. B}\ }\textbf {\bibinfo {volume} {73}},\
  \bibinfo {pages} {094423} (\bibinfo {year} {2006})}\BibitemShut {NoStop}%
\bibitem [{\citenamefont {White}(1992)}]{white_density_1992}%
  \BibitemOpen
  \bibfield  {author} {\bibinfo {author} {\bibfnamefont {S.~R.}\ \bibnamefont
  {White}},\ }\href {https://link.aps.org/doi/10.1103/PhysRevLett.69.2863}
  {\bibfield  {journal} {\bibinfo  {journal} {Phys. Rev. Lett.}\ }\textbf
  {\bibinfo {volume} {69}},\ \bibinfo {pages} {2863} (\bibinfo {year}
  {1992})}\BibitemShut {NoStop}%
\bibitem [{\citenamefont {Haegeman}\ \emph {et~al.}(2016)\citenamefont
  {Haegeman}, \citenamefont {Lubich}, \citenamefont {Oseledets}, \citenamefont
  {Vandereycken},\ and\ \citenamefont {Verstraete}}]{haegeman_unifying_2016}%
  \BibitemOpen
  \bibfield  {author} {\bibinfo {author} {\bibfnamefont {J.}~\bibnamefont
  {Haegeman}}, \bibinfo {author} {\bibfnamefont {C.}~\bibnamefont {Lubich}},
  \bibinfo {author} {\bibfnamefont {I.}~\bibnamefont {Oseledets}}, \bibinfo
  {author} {\bibfnamefont {B.}~\bibnamefont {Vandereycken}}, \ and\ \bibinfo
  {author} {\bibfnamefont {F.}~\bibnamefont {Verstraete}},\ }\href
  {https://link.aps.org/doi/10.1103/PhysRevB.94.165116} {\bibfield  {journal}
  {\bibinfo  {journal} {Phys. Rev. B}\ }\textbf {\bibinfo {volume} {94}},\
  \bibinfo {pages} {165116} (\bibinfo {year} {2016})}\BibitemShut {NoStop}%
\bibitem [{\citenamefont {Zaletel}\ \emph {et~al.}(2015)\citenamefont
  {Zaletel}, \citenamefont {Mong}, \citenamefont {Karrasch}, \citenamefont
  {Moore},\ and\ \citenamefont {Pollmann}}]{zaletel_time-evolving_2015}%
  \BibitemOpen
  \bibfield  {author} {\bibinfo {author} {\bibfnamefont {M.~P.}\ \bibnamefont
  {Zaletel}}, \bibinfo {author} {\bibfnamefont {R.~S.~K.}\ \bibnamefont
  {Mong}}, \bibinfo {author} {\bibfnamefont {C.}~\bibnamefont {Karrasch}},
  \bibinfo {author} {\bibfnamefont {J.~E.}\ \bibnamefont {Moore}}, \ and\
  \bibinfo {author} {\bibfnamefont {F.}~\bibnamefont {Pollmann}},\ }\href
  {https://link.aps.org/doi/10.1103/PhysRevB.91.165112} {\bibfield  {journal}
  {\bibinfo  {journal} {Phys. Rev. B}\ }\textbf {\bibinfo {volume} {91}},\
  \bibinfo {pages} {165112} (\bibinfo {year} {2015})}\BibitemShut {NoStop}%
\bibitem [{\citenamefont {W\'{o}jtowicz}\ \emph {et~al.}(2019)\citenamefont
  {W\'{o}jtowicz}, \citenamefont {Elenewski}, \citenamefont {Rams},\ and\
  \citenamefont {Zwolak}}]{2019inprep}%
  \BibitemOpen
  \bibfield  {author} {\bibinfo {author} {\bibfnamefont {G.}~\bibnamefont
  {W\'{o}jtowicz}}, \bibinfo {author} {\bibfnamefont {J.}~\bibnamefont
  {Elenewski}}, \bibinfo {author} {\bibfnamefont {M.~M.}\ \bibnamefont {Rams}},
  \ and\ \bibinfo {author} {\bibfnamefont {M.}~\bibnamefont {Zwolak}},\ }\href
  {http://arxiv.org/abs/1911.09108} {\bibfield  {journal} {\bibinfo  {journal}
  {arXiv:1911.09108}\ } (\bibinfo {year} {2019})}\BibitemShut {NoStop}%
\bibitem [{\citenamefont {Gruss}\ \emph {et~al.}(2016)\citenamefont {Gruss},
  \citenamefont {Velizhanin},\ and\ \citenamefont
  {Zwolak}}]{gruss_landauers_2016}%
  \BibitemOpen
  \bibfield  {author} {\bibinfo {author} {\bibfnamefont {D.}~\bibnamefont
  {Gruss}}, \bibinfo {author} {\bibfnamefont {K.~A.}\ \bibnamefont
  {Velizhanin}}, \ and\ \bibinfo {author} {\bibfnamefont {M.}~\bibnamefont
  {Zwolak}},\ }\href
  {http://www.nature.com/srep/2016/160420/srep24514/full/srep24514.html}
  {\bibfield  {journal} {\bibinfo  {journal} {Sci. Rep.}\ }\textbf {\bibinfo
  {volume} {6}},\ \bibinfo {pages} {24514} (\bibinfo {year}
  {2016})}\BibitemShut {NoStop}%
\bibitem [{\citenamefont {Gruss}\ \emph {et~al.}(2017)\citenamefont {Gruss},
  \citenamefont {Smolyanitsky},\ and\ \citenamefont
  {Zwolak}}]{gruss_communication:_2017}%
  \BibitemOpen
  \bibfield  {author} {\bibinfo {author} {\bibfnamefont {D.}~\bibnamefont
  {Gruss}}, \bibinfo {author} {\bibfnamefont {A.}~\bibnamefont {Smolyanitsky}},
  \ and\ \bibinfo {author} {\bibfnamefont {M.}~\bibnamefont {Zwolak}},\ }\href
  {https://aip.scitation.org/doi/abs/10.1063/1.4997022} {\bibfield  {journal}
  {\bibinfo  {journal} {J. Chem. Phys.}\ }\textbf {\bibinfo {volume} {147}},\
  \bibinfo {pages} {141102} (\bibinfo {year} {2017})}\BibitemShut {NoStop}%
\bibitem [{\citenamefont {Elenewski}\ \emph {et~al.}(2017)\citenamefont
  {Elenewski}, \citenamefont {Gruss},\ and\ \citenamefont
  {Zwolak}}]{elenewski_communication:_2017}%
  \BibitemOpen
  \bibfield  {author} {\bibinfo {author} {\bibfnamefont {J.~E.}\ \bibnamefont
  {Elenewski}}, \bibinfo {author} {\bibfnamefont {D.}~\bibnamefont {Gruss}}, \
  and\ \bibinfo {author} {\bibfnamefont {M.}~\bibnamefont {Zwolak}},\ }\href
  {https://aip.scitation.org/doi/10.1063/1.5000747} {\bibfield  {journal}
  {\bibinfo  {journal} {J. Chem. Phys.}\ }\textbf {\bibinfo {volume} {147}},\
  \bibinfo {pages} {151101} (\bibinfo {year} {2017})}\BibitemShut {NoStop}%
\bibitem [{\citenamefont {Zwolak}(2008)}]{Zwolak08-2}%
  \BibitemOpen
  \bibfield  {author} {\bibinfo {author} {\bibfnamefont {M.}~\bibnamefont
  {Zwolak}},\ }\href {https://doi.org/10.1063/1.2976008} {\bibfield  {journal}
  {\bibinfo  {journal} {J. Chem. Phys.}\ }\textbf {\bibinfo {volume} {129}},\
  \bibinfo {pages} {101101} (\bibinfo {year} {2008})}\BibitemShut {NoStop}%
\bibitem [{\citenamefont {Chien}\ \emph {et~al.}(2012)\citenamefont {Chien},
  \citenamefont {Zwolak},\ and\ \citenamefont
  {Di~Ventra}}]{chien_bosonic_2012}%
  \BibitemOpen
  \bibfield  {author} {\bibinfo {author} {\bibfnamefont {C.-C.}\ \bibnamefont
  {Chien}}, \bibinfo {author} {\bibfnamefont {M.}~\bibnamefont {Zwolak}}, \
  and\ \bibinfo {author} {\bibfnamefont {M.}~\bibnamefont {Di~Ventra}},\ }\href
  {\doibase 10.1103/PhysRevA.85.041601} {\bibfield  {journal} {\bibinfo
  {journal} {Phys. Rev. A}\ }\textbf {\bibinfo {volume} {85}},\ \bibinfo
  {pages} {041601} (\bibinfo {year} {2012})}\BibitemShut {NoStop}%
\bibitem [{\citenamefont {Anderson}(1961)}]{anderson_localized_1961}%
  \BibitemOpen
  \bibfield  {author} {\bibinfo {author} {\bibfnamefont {P.~W.}\ \bibnamefont
  {Anderson}},\ }\href {https://link.aps.org/doi/10.1103/PhysRev.124.41}
  {\bibfield  {journal} {\bibinfo  {journal} {Physical Review}\ }\textbf
  {\bibinfo {volume} {124}},\ \bibinfo {pages} {41} (\bibinfo {year}
  {1961})}\BibitemShut {NoStop}%
\bibitem [{\citenamefont {Bauer}\ \emph {et~al.}(2013)\citenamefont {Bauer},
  \citenamefont {Heyder}, \citenamefont {Schubert}, \citenamefont {Borowsky},
  \citenamefont {Taubert}, \citenamefont {Bruognolo}, \citenamefont {Schuh},
  \citenamefont {Wegscheider}, \citenamefont {von Delft},\ and\ \citenamefont
  {Ludwig}}]{bauer_microscopic_2013}%
  \BibitemOpen
  \bibfield  {author} {\bibinfo {author} {\bibfnamefont {F.}~\bibnamefont
  {Bauer}}, \bibinfo {author} {\bibfnamefont {J.}~\bibnamefont {Heyder}},
  \bibinfo {author} {\bibfnamefont {E.}~\bibnamefont {Schubert}}, \bibinfo
  {author} {\bibfnamefont {D.}~\bibnamefont {Borowsky}}, \bibinfo {author}
  {\bibfnamefont {D.}~\bibnamefont {Taubert}}, \bibinfo {author} {\bibfnamefont
  {B.}~\bibnamefont {Bruognolo}}, \bibinfo {author} {\bibfnamefont
  {D.}~\bibnamefont {Schuh}}, \bibinfo {author} {\bibfnamefont
  {W.}~\bibnamefont {Wegscheider}}, \bibinfo {author} {\bibfnamefont
  {J.}~\bibnamefont {von Delft}}, \ and\ \bibinfo {author} {\bibfnamefont
  {S.}~\bibnamefont {Ludwig}},\ }\href
  {https://www.nature.com/articles/nature12421} {\bibfield  {journal} {\bibinfo
   {journal} {Nature}\ }\textbf {\bibinfo {volume} {501}},\ \bibinfo {pages}
  {73} (\bibinfo {year} {2013})}\BibitemShut {NoStop}%
\bibitem [{\citenamefont {Iqbal}\ \emph {et~al.}(2013)\citenamefont {Iqbal},
  \citenamefont {Levy}, \citenamefont {Koop}, \citenamefont {Dekker},
  \citenamefont {de~Jong}, \citenamefont {van~der Velde}, \citenamefont
  {Reuter}, \citenamefont {Wieck}, \citenamefont {Aguado}, \citenamefont
  {Meir},\ and\ \citenamefont {van~der Wal}}]{iqbal_odd_2013}%
  \BibitemOpen
  \bibfield  {author} {\bibinfo {author} {\bibfnamefont {M.~J.}\ \bibnamefont
  {Iqbal}}, \bibinfo {author} {\bibfnamefont {R.}~\bibnamefont {Levy}},
  \bibinfo {author} {\bibfnamefont {E.~J.}\ \bibnamefont {Koop}}, \bibinfo
  {author} {\bibfnamefont {J.~B.}\ \bibnamefont {Dekker}}, \bibinfo {author}
  {\bibfnamefont {J.~P.}\ \bibnamefont {de~Jong}}, \bibinfo {author}
  {\bibfnamefont {J.~H.~M.}\ \bibnamefont {van~der Velde}}, \bibinfo {author}
  {\bibfnamefont {D.}~\bibnamefont {Reuter}}, \bibinfo {author} {\bibfnamefont
  {A.~D.}\ \bibnamefont {Wieck}}, \bibinfo {author} {\bibfnamefont
  {R.}~\bibnamefont {Aguado}}, \bibinfo {author} {\bibfnamefont
  {Y.}~\bibnamefont {Meir}}, \ and\ \bibinfo {author} {\bibfnamefont {C.~H.}\
  \bibnamefont {van~der Wal}},\ }\href
  {https://www.nature.com/articles/nature12491} {\bibfield  {journal} {\bibinfo
   {journal} {Nature}\ }\textbf {\bibinfo {volume} {501}},\ \bibinfo {pages}
  {79} (\bibinfo {year} {2013})}\BibitemShut {NoStop}%
\bibitem [{\citenamefont {Shi}\ \emph {et~al.}(2006)\citenamefont {Shi},
  \citenamefont {Duan},\ and\ \citenamefont {Vidal}}]{Vidal_tree_2006}%
  \BibitemOpen
  \bibfield  {author} {\bibinfo {author} {\bibfnamefont {Y.-Y.}\ \bibnamefont
  {Shi}}, \bibinfo {author} {\bibfnamefont {L.-M.}\ \bibnamefont {Duan}}, \
  and\ \bibinfo {author} {\bibfnamefont {G.}~\bibnamefont {Vidal}},\ }\href
  {\doibase 10.1103/PhysRevA.74.022320} {\bibfield  {journal} {\bibinfo
  {journal} {Phys. Rev. A}\ }\textbf {\bibinfo {volume} {74}},\ \bibinfo
  {pages} {022320} (\bibinfo {year} {2006})}\BibitemShut {NoStop}%
\bibitem [{\citenamefont {Murg}\ \emph {et~al.}(2010)\citenamefont {Murg},
  \citenamefont {Verstraete}, \citenamefont {Legeza},\ and\ \citenamefont
  {Noack}}]{Murg_tree_2010}%
  \BibitemOpen
  \bibfield  {author} {\bibinfo {author} {\bibfnamefont {V.}~\bibnamefont
  {Murg}}, \bibinfo {author} {\bibfnamefont {F.}~\bibnamefont {Verstraete}},
  \bibinfo {author} {\bibfnamefont {{\"O}.}~\bibnamefont {Legeza}}, \ and\
  \bibinfo {author} {\bibfnamefont {R.~M.}\ \bibnamefont {Noack}},\ }\href
  {\doibase 10.1103/PhysRevB.82.205105} {\bibfield  {journal} {\bibinfo
  {journal} {Phys. Rev. B}\ }\textbf {\bibinfo {volume} {82}},\ \bibinfo
  {pages} {205105} (\bibinfo {year} {2010})}\BibitemShut {NoStop}%
\bibitem [{\citenamefont {Bauernfeind}\ \emph {et~al.}(2017)\citenamefont
  {Bauernfeind}, \citenamefont {Zingl}, \citenamefont {Triebl}, \citenamefont
  {Aichhorn},\ and\ \citenamefont {Evertz}}]{bauernfeind_fork_2017}%
  \BibitemOpen
  \bibfield  {author} {\bibinfo {author} {\bibfnamefont {D.}~\bibnamefont
  {Bauernfeind}}, \bibinfo {author} {\bibfnamefont {M.}~\bibnamefont {Zingl}},
  \bibinfo {author} {\bibfnamefont {R.}~\bibnamefont {Triebl}}, \bibinfo
  {author} {\bibfnamefont {M.}~\bibnamefont {Aichhorn}}, \ and\ \bibinfo
  {author} {\bibfnamefont {H.~G.}\ \bibnamefont {Evertz}},\ }\href
  {https://link.aps.org/doi/10.1103/PhysRevX.7.031013} {\bibfield  {journal}
  {\bibinfo  {journal} {Phys. Rev. X}\ }\textbf {\bibinfo {volume} {7}},\
  \bibinfo {pages} {031013} (\bibinfo {year} {2017})}\BibitemShut {NoStop}%
\bibitem [{\citenamefont {Dorda}\ \emph {et~al.}(2015)\citenamefont {Dorda},
  \citenamefont {Ganahl}, \citenamefont {Evertz}, \citenamefont {von~der
  Linden},\ and\ \citenamefont {Arrigoni}}]{Dora_master_2015}%
  \BibitemOpen
  \bibfield  {author} {\bibinfo {author} {\bibfnamefont {A.}~\bibnamefont
  {Dorda}}, \bibinfo {author} {\bibfnamefont {M.}~\bibnamefont {Ganahl}},
  \bibinfo {author} {\bibfnamefont {H.~G.}\ \bibnamefont {Evertz}}, \bibinfo
  {author} {\bibfnamefont {W.}~\bibnamefont {von~der Linden}}, \ and\ \bibinfo
  {author} {\bibfnamefont {E.}~\bibnamefont {Arrigoni}},\ }\href {\doibase
  10.1103/PhysRevB.92.125145} {\bibfield  {journal} {\bibinfo  {journal} {Phys.
  Rev. B}\ }\textbf {\bibinfo {volume} {92}},\ \bibinfo {pages} {125145}
  (\bibinfo {year} {2015})}\BibitemShut {NoStop}%
\bibitem [{\citenamefont {Schwarz}\ \emph {et~al.}(2018)\citenamefont
  {Schwarz}, \citenamefont {Weymann}, \citenamefont {von Delft},\ and\
  \citenamefont {Weichselbaum}}]{schwarz_nonequilibrium_2018}%
  \BibitemOpen
  \bibfield  {author} {\bibinfo {author} {\bibfnamefont {F.}~\bibnamefont
  {Schwarz}}, \bibinfo {author} {\bibfnamefont {I.}~\bibnamefont {Weymann}},
  \bibinfo {author} {\bibfnamefont {J.}~\bibnamefont {von Delft}}, \ and\
  \bibinfo {author} {\bibfnamefont {A.}~\bibnamefont {Weichselbaum}},\ }\href
  {https://link.aps.org/doi/10.1103/PhysRevLett.121.137702} {\bibfield
  {journal} {\bibinfo  {journal} {Phys. Rev. Lett.}\ }\textbf {\bibinfo
  {volume} {121}},\ \bibinfo {pages} {137702} (\bibinfo {year}
  {2018})}\BibitemShut {NoStop}%
\bibitem [{\citenamefont {Fugger}\ \emph {et~al.}(2018)\citenamefont {Fugger},
  \citenamefont {Dorda}, \citenamefont {Schwarz}, \citenamefont {von Delft},\
  and\ \citenamefont {Arrigoni}}]{Fugger_kondo_magnetic_2018}%
  \BibitemOpen
  \bibfield  {author} {\bibinfo {author} {\bibfnamefont {D.~M.}\ \bibnamefont
  {Fugger}}, \bibinfo {author} {\bibfnamefont {A.}~\bibnamefont {Dorda}},
  \bibinfo {author} {\bibfnamefont {F.}~\bibnamefont {Schwarz}}, \bibinfo
  {author} {\bibfnamefont {J.}~\bibnamefont {von Delft}}, \ and\ \bibinfo
  {author} {\bibfnamefont {E.}~\bibnamefont {Arrigoni}},\ }\href {\doibase
  10.1088/1367-2630/aa9fdc} {\bibfield  {journal} {\bibinfo  {journal} {New
  Journal of Physics}\ }\textbf {\bibinfo {volume} {20}},\ \bibinfo {pages}
  {013030} (\bibinfo {year} {2018})}\BibitemShut {NoStop}%
\bibitem [{\citenamefont {Krumnow}\ \emph {et~al.}(2016)\citenamefont
  {Krumnow}, \citenamefont {Veis}, \citenamefont {Legeza},\ and\ \citenamefont
  {Eisert}}]{krumnow_fermionic_2016}%
  \BibitemOpen
  \bibfield  {author} {\bibinfo {author} {\bibfnamefont {C.}~\bibnamefont
  {Krumnow}}, \bibinfo {author} {\bibfnamefont {L.}~\bibnamefont {Veis}},
  \bibinfo {author} {\bibfnamefont {{\"O}.}~\bibnamefont {Legeza}}, \ and\
  \bibinfo {author} {\bibfnamefont {J.}~\bibnamefont {Eisert}},\ }\href
  {\doibase 10.1103/PhysRevLett.117.210402} {\bibfield  {journal} {\bibinfo
  {journal} {Phys. Rev. Lett.}\ }\textbf {\bibinfo {volume} {117}},\ \bibinfo
  {pages} {210402} (\bibinfo {year} {2016})}\BibitemShut {NoStop}%
\bibitem [{\citenamefont {Krumnow}\ \emph {et~al.}(2019)\citenamefont
  {Krumnow}, \citenamefont {Eisert},\ and\ \citenamefont {Legeza}}]{Jens2019}%
  \BibitemOpen
  \bibfield  {author} {\bibinfo {author} {\bibfnamefont {C.}~\bibnamefont
  {Krumnow}}, \bibinfo {author} {\bibfnamefont {J.}~\bibnamefont {Eisert}}, \
  and\ \bibinfo {author} {\bibfnamefont {{\"O}.}~\bibnamefont {Legeza}},\
  }\href {http://arxiv.org/abs/1904.11999} {\bibfield  {journal} {\bibinfo
  {journal} {arXiv:1904.11999}\ } (\bibinfo {year} {2019})}\BibitemShut
  {NoStop}%
\bibitem [{\citenamefont {Jauho}\ \emph {et~al.}(1994)\citenamefont {Jauho},
  \citenamefont {Wingreen},\ and\ \citenamefont
  {Meir}}]{jauho_time-dependent_1994}%
  \BibitemOpen
  \bibfield  {author} {\bibinfo {author} {\bibfnamefont {A.-P.}\ \bibnamefont
  {Jauho}}, \bibinfo {author} {\bibfnamefont {N.~S.}\ \bibnamefont {Wingreen}},
  \ and\ \bibinfo {author} {\bibfnamefont {Y.}~\bibnamefont {Meir}},\ }\href
  {http://link.aps.org/doi/10.1103/PhysRevB.50.5528} {\bibfield  {journal}
  {\bibinfo  {journal} {Phys. Rev. B}\ }\textbf {\bibinfo {volume} {50}},\
  \bibinfo {pages} {5528} (\bibinfo {year} {1994})}\BibitemShut {NoStop}%
\bibitem [{\citenamefont {Caroli}\ \emph {et~al.}(1971)\citenamefont {Caroli},
  \citenamefont {Combescot}, \citenamefont {Nozieres},\ and\ \citenamefont
  {Saint-James}}]{caroli_direct_1971}%
  \BibitemOpen
  \bibfield  {author} {\bibinfo {author} {\bibfnamefont {C.}~\bibnamefont
  {Caroli}}, \bibinfo {author} {\bibfnamefont {R.}~\bibnamefont {Combescot}},
  \bibinfo {author} {\bibfnamefont {P.}~\bibnamefont {Nozieres}}, \ and\
  \bibinfo {author} {\bibfnamefont {D.}~\bibnamefont {Saint-James}},\ }\href
  {http://stacks.iop.org/0022-3719/4/i=8/a=018} {\bibfield  {journal} {\bibinfo
   {journal} {J. Phys. C: Solid State Phys.}\ }\textbf {\bibinfo {volume}
  {4}},\ \bibinfo {pages} {916} (\bibinfo {year} {1971})}\BibitemShut {NoStop}%
\bibitem [{\citenamefont {Wilson}(1975)}]{Wilson75-1}%
  \BibitemOpen
  \bibfield  {author} {\bibinfo {author} {\bibfnamefont {K.~G.}\ \bibnamefont
  {Wilson}},\ }\href {\doibase 10.1103/RevModPhys.47.773} {\bibfield  {journal}
  {\bibinfo  {journal} {Rev. Mod. Phys.}\ }\textbf {\bibinfo {volume} {47}},\
  \bibinfo {pages} {773} (\bibinfo {year} {1975})}\BibitemShut {NoStop}%
\bibitem [{\citenamefont {Bulla}\ \emph {et~al.}(2008)\citenamefont {Bulla},
  \citenamefont {Costi},\ and\ \citenamefont
  {Pruschke}}]{bulla_numerical_2008}%
  \BibitemOpen
  \bibfield  {author} {\bibinfo {author} {\bibfnamefont {R.}~\bibnamefont
  {Bulla}}, \bibinfo {author} {\bibfnamefont {T.~A.}\ \bibnamefont {Costi}}, \
  and\ \bibinfo {author} {\bibfnamefont {T.}~\bibnamefont {Pruschke}},\ }\href
  {https://link.aps.org/doi/10.1103/RevModPhys.80.395} {\bibfield  {journal}
  {\bibinfo  {journal} {Rev. Mod. Phys.}\ }\textbf {\bibinfo {volume} {80}},\
  \bibinfo {pages} {395} (\bibinfo {year} {2008})}\BibitemShut {NoStop}%
\bibitem [{\citenamefont {Vidal}(2003)}]{vidal_efficient_2003}%
  \BibitemOpen
  \bibfield  {author} {\bibinfo {author} {\bibfnamefont {G.}~\bibnamefont
  {Vidal}},\ }\href {https://link.aps.org/doi/10.1103/PhysRevLett.91.147902}
  {\bibfield  {journal} {\bibinfo  {journal} {Phys. Rev. Lett.}\ }\textbf
  {\bibinfo {volume} {91}},\ \bibinfo {pages} {147902} (\bibinfo {year}
  {2003})}\BibitemShut {NoStop}%
\bibitem [{\citenamefont {Vidal}(2004)}]{vidal_efficient_2004}%
  \BibitemOpen
  \bibfield  {author} {\bibinfo {author} {\bibfnamefont {G.}~\bibnamefont
  {Vidal}},\ }\href {https://link.aps.org/doi/10.1103/PhysRevLett.93.040502}
  {\bibfield  {journal} {\bibinfo  {journal} {Phys. Rev. Lett.}\ }\textbf
  {\bibinfo {volume} {93}},\ \bibinfo {pages} {040502} (\bibinfo {year}
  {2004})}\BibitemShut {NoStop}%
\bibitem [{\citenamefont
  {Schmitteckert}(2004)}]{schmitteckert_nonequilibrium_2004}%
  \BibitemOpen
  \bibfield  {author} {\bibinfo {author} {\bibfnamefont {P.}~\bibnamefont
  {Schmitteckert}},\ }\href
  {https://link.aps.org/doi/10.1103/PhysRevB.70.121302} {\bibfield  {journal}
  {\bibinfo  {journal} {Phys. Rev. B}\ }\textbf {\bibinfo {volume} {70}},\
  \bibinfo {pages} {121302} (\bibinfo {year} {2004})}\BibitemShut {NoStop}%
\bibitem [{\citenamefont {Haegeman}\ \emph {et~al.}(2011)\citenamefont
  {Haegeman}, \citenamefont {Cirac}, \citenamefont {Osborne}, \citenamefont
  {Pizorn}, \citenamefont {Verschelde},\ and\ \citenamefont
  {Verstraete}}]{haegeman_time-dependent_2011}%
  \BibitemOpen
  \bibfield  {author} {\bibinfo {author} {\bibfnamefont {J.}~\bibnamefont
  {Haegeman}}, \bibinfo {author} {\bibfnamefont {J.~I.}\ \bibnamefont {Cirac}},
  \bibinfo {author} {\bibfnamefont {T.~J.}\ \bibnamefont {Osborne}}, \bibinfo
  {author} {\bibfnamefont {I.}~\bibnamefont {Pizorn}}, \bibinfo {author}
  {\bibfnamefont {H.}~\bibnamefont {Verschelde}}, \ and\ \bibinfo {author}
  {\bibfnamefont {F.}~\bibnamefont {Verstraete}},\ }\href
  {https://link.aps.org/doi/10.1103/PhysRevLett.107.070601} {\bibfield
  {journal} {\bibinfo  {journal} {Phys. Rev. Lett.}\ }\textbf {\bibinfo
  {volume} {107}},\ \bibinfo {pages} {070601} (\bibinfo {year}
  {2011})}\BibitemShut {NoStop}%
\bibitem [{\citenamefont {Niesen}\ and\ \citenamefont
  {Wright}(2012)}]{Niesen_Krylov}%
  \BibitemOpen
  \bibfield  {author} {\bibinfo {author} {\bibfnamefont {J.}~\bibnamefont
  {Niesen}}\ and\ \bibinfo {author} {\bibfnamefont {W.~M.}\ \bibnamefont
  {Wright}},\ }\href {\doibase 10.1145/2168773.2168781} {\bibfield  {journal}
  {\bibinfo  {journal} {ACM Trans. Math. Softw.}\ }\textbf {\bibinfo {volume}
  {38}},\ \bibinfo {pages} {22:1} (\bibinfo {year} {2012})}\BibitemShut
  {NoStop}%
\bibitem [{\citenamefont {Zwolak}(2018)}]{zwolak_communication:_2018}%
  \BibitemOpen
  \bibfield  {author} {\bibinfo {author} {\bibfnamefont {M.}~\bibnamefont
  {Zwolak}},\ }\href {https://aip.scitation.org/doi/10.1063/1.5061759}
  {\bibfield  {journal} {\bibinfo  {journal} {J. Chem. Phys.}\ }\textbf
  {\bibinfo {volume} {149}},\ \bibinfo {pages} {241102} (\bibinfo {year}
  {2018})}\BibitemShut {NoStop}%
\bibitem [{\citenamefont {Meyer}\ \emph {et~al.}(1990)\citenamefont {Meyer},
  \citenamefont {Manthe},\ and\ \citenamefont
  {Cederbaum}}]{meyer_multi-configurational_1990}%
  \BibitemOpen
  \bibfield  {author} {\bibinfo {author} {\bibfnamefont {H.~D.}\ \bibnamefont
  {Meyer}}, \bibinfo {author} {\bibfnamefont {U.}~\bibnamefont {Manthe}}, \
  and\ \bibinfo {author} {\bibfnamefont {L.~S.}\ \bibnamefont {Cederbaum}},\
  }\href {http://www.sciencedirect.com/science/article/pii/000926149087014I}
  {\bibfield  {journal} {\bibinfo  {journal} {Chem. Phys. Lett.}\ }\textbf
  {\bibinfo {volume} {165}},\ \bibinfo {pages} {73} (\bibinfo {year}
  {1990})}\BibitemShut {NoStop}%
\bibitem [{\citenamefont {Beck}\ \emph {et~al.}(2000)\citenamefont {Beck},
  \citenamefont {Jäckle}, \citenamefont {Worth},\ and\ \citenamefont
  {Meyer}}]{beck_multiconfiguration_2000}%
  \BibitemOpen
  \bibfield  {author} {\bibinfo {author} {\bibfnamefont {M.~H.}\ \bibnamefont
  {Beck}}, \bibinfo {author} {\bibfnamefont {A.}~\bibnamefont {Jäckle}},
  \bibinfo {author} {\bibfnamefont {G.~A.}\ \bibnamefont {Worth}}, \ and\
  \bibinfo {author} {\bibfnamefont {H.~D.}\ \bibnamefont {Meyer}},\ }\href
  {http://www.sciencedirect.com/science/article/pii/S0370157399000472}
  {\bibfield  {journal} {\bibinfo  {journal} {Phys. Rep.}\ }\textbf {\bibinfo
  {volume} {324}},\ \bibinfo {pages} {1} (\bibinfo {year} {2000})}\BibitemShut
  {NoStop}%
\bibitem [{\citenamefont {Kloss}\ \emph {et~al.}(2019)\citenamefont {Kloss},
  \citenamefont {Reichman},\ and\ \citenamefont
  {Tempelaar}}]{kloss_multi-set_2018}%
  \BibitemOpen
  \bibfield  {author} {\bibinfo {author} {\bibfnamefont {B.}~\bibnamefont
  {Kloss}}, \bibinfo {author} {\bibfnamefont {D.~R.}\ \bibnamefont {Reichman}},
  \ and\ \bibinfo {author} {\bibfnamefont {R.}~\bibnamefont {Tempelaar}},\
  }\href {\doibase 10.1103/PhysRevLett.123.126601} {\bibfield  {journal}
  {\bibinfo  {journal} {Phys. Rev. Lett.}\ }\textbf {\bibinfo {volume} {123}},\
  \bibinfo {pages} {126601} (\bibinfo {year} {2019})}\BibitemShut {NoStop}%
\end{thebibliography}
\end{document}